\documentclass{article}

\usepackage{graphicx} % Required for inserting images
\usepackage[round]{natbib}
\usepackage{amsmath}
\usepackage{amssymb}
\usepackage{mathtools}
\usepackage{dsfont}
\usepackage{pdflscape}
\usepackage{booktabs}
\usepackage[table]{xcolor}
\usepackage{appendix}
\usepackage[parfill]{parskip}
\usepackage[left=2cm, right=2cm, top=2cm, bottom=3cm]{geometry}
\usepackage{orcidlink}
\usepackage{subcaption}
\usepackage{hyperref}
\usepackage{longtable}
\usepackage{colortbl}
\usepackage{multicol}
\usepackage{multirow}
\usepackage{adjustbox}
\usepackage[section]{placeins}
\usepackage{tikz}
\usetikzlibrary{decorations.pathreplacing,arrows.meta,calc}
\usetikzlibrary{arrows.meta, positioning, calc, decorations.pathreplacing}

\usepackage{pgfplots}
\pgfplotsset{compat=1.18}

\usepackage{nicefrac}

\definecolor{histblue}{RGB}{55,115,190}
\definecolor{adaptorange}{RGB}{230,145,45}
\definecolor{todaygreen}{RGB}{70,145,90}
\definecolor{softblue}{RGB}{225,238,255}
\definecolor{softorange}{RGB}{255,238,220}
\definecolor{softgreen}{RGB}{232,246,235}

\usepackage[ruled,vlined, linesnumbered]{algorithm2e}
\SetKwRepeat{Do}{do}{while}
\SetKw{Return}{Return}
\SetKwComment{Comment}{\#{} }{}
\SetKwInput{KwStore}{Store}

\newcommand{\Hc}{\ensuremath{\mathcal{H}}}
\newcommand{\Oc}{\ensuremath{\mathcal{O}}}
\newcommand{\Nd}{\ensuremath{\mathds{N}}}
\newcommand{\mat}[1]{\ensuremath{\boldsymbol{#1}}}

\title{Improving Algorithmic Trading for BESS on Intraday Electricity Markets through Distributional Price Forecasts}
\author{
    Simon Hirsch\textsuperscript{1,2,\orcidlink{0009-0008-1409-9677}} \and 
    David Schaurecker\textsuperscript{3, \orcidlink{0000-0001-6746-6452}} \and 
    David Wozabal\textsuperscript{4,\orcidlink{0000-0002-6656-9529}} 
}
\date{%
    \small\textsuperscript{1}House of Energy Markets \& Finance (HEMF), University of Duisburg-Essen, Germany \\
    \small\textsuperscript{2}Statkraft Trading GmbH, Germany \\
    \small\textsuperscript{3}Eidgenössische Technische Hochschule (ETH) Zürich, Switzerland \\
    \small\textsuperscript{4}Operations Analytics Department, Vrije Universiteit Amsterdam, The Netherlands \\ [1em]
}

\begin{document}
\maketitle
%\tableofcontents

\begin{abstract}
    %Grid-scale battery energy storage systems (BESS) become increasingly important on short-term energy markets to balance fluctuations of variable renewable energy generation. Storage assets can exploit arbitrage opportunities through buying (and storing) and selling (and discharging) electricity. The continuous intraday market is characterized by high volatility and operational intricacies such as continuous limit order book trading, a special, and parallel time structure and a strong physical component. These constraints are seldom reflected in the literature on intraday price forecasting which focuses on the last three hours of trading and volume-weighted prices. We develop an integration of distributional forecasts into BESS intraday optimization. Our approach overcomes previous weaknesses by defining a novel target variable based on the limit order book and its dynamics. Our forecast is therefore tailored to inform the optimization about future profit opportunities while maintaining efficiency and avoiding the computational burden of full stochastic optimization. This improves profits in a realistic high-frequency simulation setting for the years 2023 and 2024 by up to 42\%. Based on these results, we conduct an analysis of the relationship between forecast accuracy, stability, and financial performance, establishing a clear inverse relationship. Furthermore, we find evidence that the size and duration of BESS influence both general profitability and the benefit of integrating forecasts.
    % 204 words -> more now

    Grid-scale battery energy storage systems (BESS) are becoming increasingly important in short-term energy markets, where they help balance fluctuations in variable renewable generation. In many countries, the continuous intraday market is the most important wholesale market for BESS to monetize their flexibility. In this paper, we propose an extension to the myopic rolling intrinsic policy that incorporates intraday price forecasts tailored specifically to the BESS trading problem. While a large literature addresses intraday price forecasting, most existing work focuses on volume-weighted prices during the last hours of trading and consequently fails to capture the features most relevant to BESS trading. Our approach addresses this gap by introducing a novel target variable derived from the limit order book and its dynamics. The resulting forecasts are designed to inform the optimization about future profit opportunities while preserving computational efficiency and avoiding the burden of full stochastic optimization. In a realistic high-frequency simulation setting covering 2023 and 2024, our method improves profits by up to 42\% over the naive rolling intrinsic formulation. We further analyze the relationship between forecast accuracy, stability, and financial performance, establishing a clear inverse relationship between accuracy and stability. Finally, we find evidence that BESS size and duration jointly influence both overall profitability and the marginal benefit of incorporating forecasts.
\end{abstract}

%%%%%%%%%%%%%%%%%%%%%%%%%%%%%%%
%%% Section 1: Introduction %%%
%%%%%%%%%%%%%%%%%%%%%%%%%%%%%%%
\section{Introduction}\label{sec:introduction}
Climate change is the pressing challenge of the 21st century. Widespread electrification and the deployment of renewable energy are seen as crucial parts of the solution. However, electricity is a unique commodity, as supply and demand must be balanced at all times. The inherent uncertainty of variable renewable energy production has increased the importance of short-term electricity markets to react to changes in the fluctuating output of renewable energy sources. Concurrently, grid-scale battery energy storage systems (BESS) have become vital for stabilizing the grid while also shifting supply from hours of abundance to hours of scarcity for financial profit. Historically, BESS have primarily operated in the ancillary services market. However, increased volatility in wholesale electricity prices has expanded opportunities for arbitrage profits through buying (and storing) and selling (and discharging) electricity \citep{gottfried2026future}. Against this background, this paper develops forecast-based trading strategies for BESS on the European continuous intraday electricity markets, which offer the last opportunity to trade. %The results of our realistic backtesting show that the inclusion of forecasts can improve financial performance by up to 42\%{} depending on the BESS' size and duration. 
Our strategy matches the millisecond pace of the continuous intraday market, re-optimizing decisions in real time while accurately modeling complex liquidity and market constraints dictated by the full limit order book. 

We build upon two extensive streams of academic literature: short-term electricity price forecasting and BESS operational optimization, which we review in the following, starting with the latter. The complexity and continuous nature of storage optimization on the intraday market induce optimization problems with a large number of decision variables and stages. 

\cite{deschatre2024common, cogneville2024battery, seifert2024coordinated, lohndorf2023value, miskiw2025coordinated, kraft_2023} propose stochastic optimization approaches for the intraday market and across wholesale and ancillary services markets. Due to the complexity of the problem, these approaches are necessarily stylized and do not directly yield implementable trading policies of the intraday market.

Therefore, many authors propose that myopic or rolling intrinsic optimization strategies are popular in \citep{gottfried2026efficient, semmelmann2025algorithm, hornek2025value, oeltz2025rolling, schaurecker2025maximizing}.

Furthermore, there is an increase in  stylized BESS optimization approaches that are used for the monetary evaluation of forecasts \citep{maciejowska2026statistical, serafin2025data, serafin2025loss}. This strand of research is interesting in the context of this paper since it analyzes the relationship between different loss functions and evaluation metrics for forecasts and day-ahead BESS profits.

Turning to the literature on electricity price forecasting (EPF), we refer the reader to the classic reviews of \cite{weron2014electricity} and \cite{nowotarski2018recent} for an introduction and to \cite{yu2026deep,o2025evolution, thakare2023forecasting} for an overview of (probabilistic) forecasting for intraday markets. 

Table~\ref{tab:literature_overview} surveys the literature on intraday electricity price forecasting, demonstrating two limitations of contemporary approaches: First, there is a strong focus on the last three hours of trading. However, decisions in storage optimization are made throughout the day, and thus forecasts issued with short lead times are insufficient. Notable exemptions are the works by \cite{puc2024corrected}, \cite{nickelsen2025bayesian} and \cite{hirsch2024multivariate}. Secondly, there is a strong emphasis on (probabilistic) forecasting of volume-weighted average prices and the intraday indices, which fail to reflect the available options for traders in the limit order book. Forecasting approaches based on data from the limit order book are sparse \citep[see][]{grindel2022forecasting, yu2025orderfusion, yu2025orderbook}, and to the best of the authors' knowledge, no attempts have been made to forecast the order book or its derivative variables directly.

Taking into account these challenges, we introduce a novel approach to integrate distributional forecasts into a rolling intrinsic optimization strategy for intraday battery trading. Our approach is based on the orderbook-based roling intrinsic trading strategy proposed by \cite{schaurecker2025maximizing} and considers the intricacies of the intraday market, such as the continuous trading mechanism, the full dynamics of the limit order book, and the operational constraints of the battery.  

The contributions of our paper are threefold:
\begin{itemize}
    \item We are the first to integrate distributional forecasts in a rolling high-frequency battery trading strategy on the intraday market. Our framework is general and can be applied to most European intraday markets and potentially different forecasting approaches.
    
    \item We discuss the challenges associated with the integration of forecasts in a high-frequency optimization and, inspired by \cite{janke2019forecasting,berrisch2023high}, propose a novel approach to intraday electricity price forecasting by taking the empirical distribution (ECDF) of mid prices for the rest of the trading session as the target variable. 
    
    \item We provide an extensive empirical analysis of the value of forecasts and examine the relationship between forecasting performance and optimization performance. We find that forecasts can significantly improve the performance of the rolling intrinsic optimization and that the relationship between forecasting performance and optimization performance is not straightforward, which highlights the importance of using appropriate evaluation metrics for both forecasting and optimization.
\end{itemize}
In a numerical study, we find that the use of forecasts significantly improves the performance of the rolling intrinsic and improves returns by up to 27\% for 2h batteries and 42\% for 4h batteries. Gains increase with battery size and duration. We observe that the forecast-based policy trades closer to the delivery of power and find a strong negative correlation between error metrics, forecast stability, and daily profits, indicating that predictive accuracy is a driving factor in returns.

The remainder of this paper is organized as follows. In the following Section~\ref{sec:limit-order-book} we elaborate on the structure of the intraday market as a limit order book based continuous market. Section~\ref{sec:optimization-with-forecasts} discusses the optimization problem and the integration of forecasts into the rolling intrinsic optimization. In Section~\ref{sec:forecasting} we discuss the forecasting approach, especially the choice of the target variable and the forecasting models. Section~\ref{sec:forecasting-results} presents our forecasting results. Finally, in Section~\ref{sec:simulation-results} we present our results on the value of forecasts for storage optimization on the intraday market and conclude in Section~\ref{sec:conclusion}.

\begin{table}[t]
    \resizebox{\linewidth}{!}{%
        \begin{tabular}{llllll}
        \toprule
        Reference                               & Target        & First issue $t$ / Updates    & Training \& Tuning      &   Evaluation     &   Point / Prob. EPF\\
        \midrule
        \cite{grindel2022forecasting}           & ID3           & ID3 / No      & 2019-01 -- 2019-05    &   2019-05 -- 2019-07      & Point \\
        \cite{narajewski2020econometric}        & ID3           & ID3 / No     & 2015-01 -- 2016-01    &   2016-01 -- 2019-10       & Point \\
        \cite{maciejowska2020pca}               & ID3           & ID3 / No      & 2015-01 -- 2017-08    &   2017-08 -- 2019-10      & Point \\
        \cite{maciejowska2024probabilistic}     & ID3           & ID3 / No      & 2015-01 -- 2017-08    &   2017-08 -- 2019-10      & Probabilistic \\
        \cite{uniejewski2019understanding}      & ID3           & ID3 / No      & 2015-01 -- 2016-01    &   2016-01 -- 2018-04      & Point \\
        %\cite{oksuz2019neural}                  & ID3           & & & \\ 
        \cite{yu2025orderfusion}                & ID3 and ID1   & ID3 / No     & 2022-01 -- 2024-07    &   2024-07 -- 2025-01   & Probabilistic\\ 
        \cite{nickelsen2025bayesian}            & ID Full       & DA / Yes, every hour           & 2020-01 -- 2022-01    &   2022-01 -- 2023-01  & Probabilistic \\ 
        %\cite{kath2021conformal}                & ID Full       & DA / No           & &  & Probabilistic \\
        \cite{narajewski2020ensemble}           & Price path    & ID3 / No      & 2015-07 -- 2016-07    &   2016-07 -- 2019-09  & Probabilistic\\
        \cite{marcjasz2023trading}              & Price path    & ID3 / No       & 2017-06 -- 2019-03    &   2018-03 -- 2019-10  & Probabilistic\\
        \cite{serafin2022trading}               & Price path    & ID3 / No       & 2017-06 -- 2019-03    &   2018-03 -- 2019-10  & Probabilistic\\ 
        \cite{hirsch2024simulation}             & Price path    & ID3 / No      & 2016-01 -- 2017-01    &   2017-01 -- 2020-07  & Probabilistic\\
        \cite{chen2025probabilistic}            & Price path    & ID3 / No      & 2017-06 -- 2019-03    &   2018-03 -- 2019-10  & Probabilistic\\
        \cite{hirsch2024multivariate}           & Price path    & DA / No            & 2020-01 -- 2022-01    &   2022-01 -- 2023-01  & Probabilistic\\
        \cite{janke2019forecasting}             & Weighted ECDF & ID3 / No      & 2019-07 -- 2018-04    &   2018-04 -- 2019-04           & In between \\
        \cite{puc2024corrected}                 & 15-min VWAP   & DA / Yes            & 2018-11 -- 2020-01    &   2020-01 -- 2021-01    & Point\\   
        \cite{scholz2021towards}                & 15-min VWAP   & ID3 / No  & \multicolumn{2}{c}{Not described. Data set from 2018.}     & Point\\
        \cite{hornek2025directional}            & Directional    & ID3 / Yes      & 30 days rolling.    &   2024-04 -- 2025-04    & Point \\
        \cite{semmelmann_cid_fc2026}            & $Q^{10}$, $Q^{90}$ of Transactions & ID3 / Yes, every hour & 2026-01 -- 2026-01 & 2026-02 -- 2026-02 & Probabilistic\\
        \bottomrule
    \end{tabular}
    }
    \caption{Overview of the literature on intraday electricity price forecasting in Germany. Issue Time refers to the first forecast issued, subsequent updates might be made at a later point. Note that some papers reserve 5-20 minutes for data processing and delays, which we abstract here. The time range of data sets is given left-closed and rounded to full months. Own compilation.}\label{tab:literature_overview}
\end{table}

%%%%%%%%%%%%%%%%%%%%%%%%%%%%%%%%%%%%%%%%%%%%%%%%%%%%%%%%%%%%%
%%% Section 2: The Continuous Intraday Electricity Market %%%
%%%%%%%%%%%%%%%%%%%%%%%%%%%%%%%%%%%%%%%%%%%%%%%%%%%%%%%%%%%%%
\section{The Continuous Intraday Electricity Market}\label{sec:limit-order-book}
\begin{figure}[t]
\definecolor{AskColor}{RGB}{229,84,81}
\definecolor{BidColor}{RGB}{76,118,131}
    \centering
    \begin{tikzpicture}
        \begin{axis}[
            axis x line=bottom,             
            xmin=0, xmax=48,
            ymin=-0.25, ymax=0.35,
            hide y axis,
            width=14cm,
            height=5cm,
            xtick={0,10,12,15,22,24,34,48},
            xticklabels={
                00:00,
                10:00,
                12:00,
                15:00,
                22:00,
                00:00,
                10:00,
                24:00
            },
            tick label style={font=\footnotesize, rotate=45, anchor=north east},
            major tick length=2mm,
        ]
        
        % --- Gray band for Intraday Continuous (IDC) trading ---
        \fill[gray!15] (axis cs:15,-0.2475) rectangle (axis cs:47,0.1);
        
        % --- Vertical dashed line separating days ---
        \draw[dashed, thin, black!60] (axis cs:24,-0.25) -- (axis cs:24,0.15);
        
        % --- Day-ahead market clearing (12:00) ---
        \draw[BidColor, thick] (axis cs:12,-0.2475) -- (axis cs:12,0.0)
            node[above, anchor=south, BidColor, font=\small] {Day-ahead};
        \draw[BidColor, thick] (axis cs:9,0.0) -- (axis cs:15,0.0);
        
        % --- First intraday auction (IDA1) and start of IDC ---
        \draw[AskColor, thick] (axis cs:15,-0.2475) -- (axis cs:15,-0.1)
            node[above, anchor=south, AskColor, font=\small] {IDA1};
        \draw[AskColor, thick] (axis cs:13.5,-0.1) -- (axis cs:16.5,-0.1);
    
        % Other intraday auctions
        \draw[AskColor, thick] (axis cs:22,-0.2475) -- (axis cs:22,-0.1)
            node[above, anchor=south, AskColor, font=\small] {IDA2};
        \draw[AskColor, thick] (axis cs:20.5,-0.1) -- (axis cs:23.5,-0.1);
    
        \draw[AskColor, thick] (axis cs:34,-0.2475) -- (axis cs:34,-0.1)
            node[above, anchor=south, AskColor, font=\small] {IDA3};
        \draw[AskColor, thick] (axis cs:32.5,-0.1) -- (axis cs:35.5,-0.1);

        % --- Label for the continuous trading period ---
        \node[anchor=center, font=\bfseries\small, text=black] at (axis cs:36.5,0.05)
            {Intraday Continuous Trading};
        %\node[anchor=north, font=\scriptsize, text=black!70] at (axis cs:31.5,-0.05)
        %    {from 15:00 Day D to 24:00 Day D+1};
        \end{axis}
    \end{tikzpicture}
    \caption{\label{fig:timing} Timing of spot markets for one day of trading. The gray area signifies the time period where products on the IDC market can be traded. The market time points are the times when the day-ahead and intraday auctions are cleared. Note that the graph does not contain the markets for the next day, which open during the depicted time period. In particular, the intraday continuous market for day $d+1$ opens at 15:00 on the second depicted day.}
\end{figure}
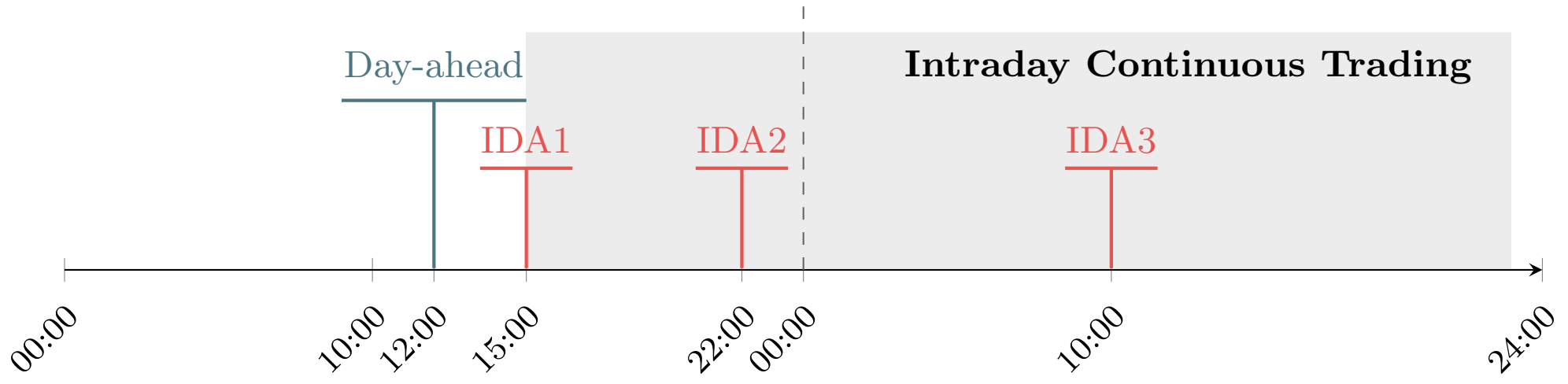
In the European internal electricity market, the continuous intraday market (IDC) is part of the spot market for electricity, which also encompasses the day-ahead market and three intraday auctions. Trading on the IDC is organized on a pan-European level and coordinated via the single intraday coupling, which enables cross-border trading where transmission capacities allow it. Unlike all other short-term electricity markets, which are organized as auctions, the IDC uses continuous trading as its settlement mechanism. Compared to auctions, this market design has the advantage of \emph{immediacy}: it allows participants to react instantaneously to new information without having to wait for auction clearing, which occurs at set discrete points in time. This comes at the cost of reduced market liquidity \cite{graf_frequent_2018}.

By its nature and due to the lack of significant system-wide storage capacities, electricity resembles more a service than a classic commodity. Furthermore, the frailty of power systems, which require perfect real-time matching of supply and demand to remain stable, necessitates short-term balancing mechanisms to address unforeseen supply and demand shocks. The main motivation for the intraday market is to provide market participants with the opportunity to adjust their positions taken in the day-ahead market close to delivery. The need for such corrections has risen in recent years with the growth of renewable energy, which leads to an increase in unpredictable fluctuations of electricity output.

The IDC market for day $d$ opens at 15:00 on day $d-1$, at which point all products for the respective next day begin to trade (see Figure~\ref{fig:timing}). The market trades futures contracts that deliver electricity for one whole hour, half an hour, or 15 minutes, meaning that $24 + 48 + 96 = 168$ products can be traded for a single day. Note that the ability to trade products for the whole day differentiates the European intraday markets from real-time markets as typically used in the US, where only the very short end of the forward curve can be traded. Currently, market participants in Germany can trade until $5$ minutes before physical delivery in their control areas and up to $30$ minutes before delivery in a combined national market \citep{kuppelwieser2023intraday}.

Each buy and sell order on the intraday market for a given product contains basic information about quantity, limit price, and validity time. A \emph{market order} is cleared immediately against the best available order in the limit order book (LOB), while a \emph{limit order} is only executed against matching orders on the other side of the market up to a specified price. If no such match is possible, the order is kept in the LOB until its \emph{end validity date} to be cleared against future orders. If the quantities of two matched orders differ, the order with the higher quantity is only partially cleared and remains in the order book with the residual quantity. Volumes are restricted to multiples of $0.1$~MW, and the price per MWh ranges from EUR $-9999$ to $9999$. Market participants can attach the usual order qualifiers such as \emph{immediate-or-cancel} (IOC) or \emph{fill-or-kill} (FOK) to their limit orders. Additionally, \emph{iceberg} orders are supported, for which only a fraction of the total order quantity is visible to other market participants, and as soon as the visible quantity is cleared, the next tranche is automatically placed in the LOB.

\begin{figure}[t]
    \centering
    \input{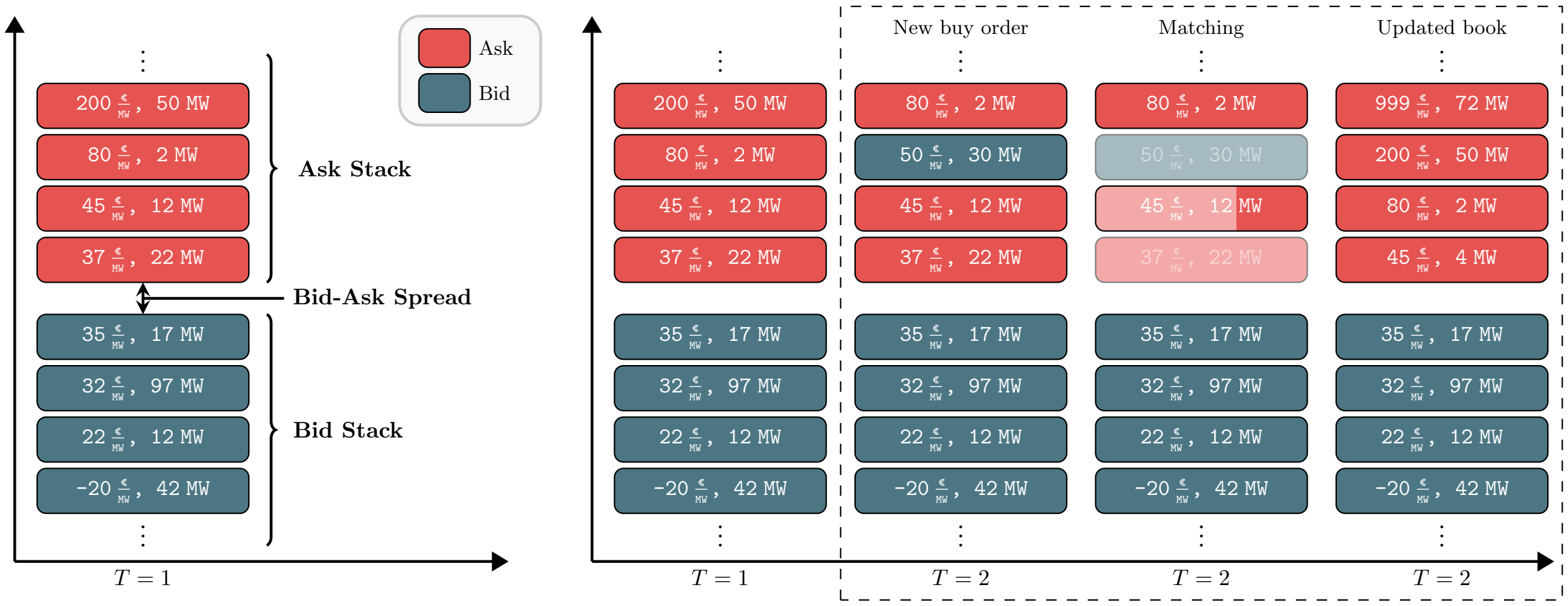}
    \caption{\label{fig:id_price} Illustration of the limit order book clearing mechanism. Figure and caption adapted from \cite{graf_frequent_2018}.}
\end{figure}

The basic mechanism of continuous trading is illustrated in Figure~\ref{fig:id_price} by a concrete example. The left panel shows the state of the order book at $T=1$, with orders sorted by limit price and organized into a \emph{bid} and an \emph{ask stack}. The state of the LOB changes with the placement or modification of another order and at the end-validity time of an active order. The lowest-priced sell order defines the \emph{best ask}, the highest-priced buy order defines the \emph{best bid}, and the difference between the two is the \emph{bid-ask spread}. The right panel illustrates the clearing of a new hypothetical buy order: a buy order at EUR 50 is added at $T=2$ and cleared against the cheapest available offers until either the full quantity is fulfilled or no offers at lower prices remain. In this example, 22~MW out of 29~MW are cleared against the sell order at EUR 37, and the remaining $29\text{~MW} - 22\text{~MW} = 7\text{~MW}$ are cleared against the sell order at 45~EUR, leaving $12\text{~MW} - 7\text{~MW} = 5\text{~MW}$ of the latter in the order book. Note that clearing is instantaneous; columns 2 and 3 in the right panel are purely illustrative and do not correspond to observable market states.

In this paper, we focus on hourly products, i.e., futures that deliver electricity in a specific hour. We denote a delivery hour as $h$, the corresponding day (or date) as $d = \text{date}(h)$, and the trading time — the point in time at which an order is placed — as $t$. We denote order prices as $P_i$ and the volumes as $V_i$, with $i$ being the unique index of the order. Alternatively, we denote $P^{[v,s]}_{h,t}$ as the price for the $v$-th MWh on side $s \in \{\text{bid, ask}\}$ at time $t$ in product $h$. For example, in Figure~\ref{fig:id_price}, $P^{10,\text{ask}}_{1} = 37$ EUR/MWh in $T=1$ for the depicted product.  

%%%%%%%%%%%%%%%%%%%%%%%%%%%%%%%%%%%%%%%%%%%%%%%%%%%%%%%%%%%%%%%%%%%%%
%%% Section 3: The Forecast-Augmented Rolling Intrinsic %%%%%%%%%%%%%
%%%%%%%%%%%%%%%%%%%%%%%%%%%%%%%%%%%%%%%%%%%%%%%%%%%%%%%%%%%%%%%%%%%%%
\section{The Forecast-Augmented Rolling Intrinsic}\label{sec:optimization-with-forecasts}
Intraday bidding optimization is a sequential decision problem under uncertainty: operators continuously rebalance positions as the limit order book evolves unpredictably. Since the number of involved random and decision variables is large, and the long continuous trading window implies a large number of decision stages, the problem is generally computationally intractable. To address these issues, the RI serves as a common heuristic for storage optimization in general and battery optimization in particular \citep{schaurecker2025maximizing, semmelmann2025algorithm}. While the RI is suboptimal, it circumvents the curse of dimensionality inherent in the aforementioned stochastic dynamic problem.

In this section, we will define an augmented version of the standard rolling intrinsic (RI) strategy that works for continuous intraday battery trading and incorporates price forecasts. We first formalize the standard rolling intrinsic policy before detailing its forecast-augmented extension motivated by a \emph{lazy clairvoyant} strategy.

%%% Section 3.1: The Rolling Intrinsic Policy
\subsection{The Rolling Intrinsic Policy}
In this section, we formalize the intrinsic value of a battery storage system in the continuous intraday market. Conceptually, the intrinsic value captures a myopic, single-period trading optimization, conditioned on an initial state-of-charge (SoC) $s_0$ and pre-existing contractual positions $f_h^0$ across all currently tradable products $h \in \mathcal{H}$.

We closely follow the formulation in \citet{schaurecker2025maximizing}. To this end, we denote by $\Oc_h^\pm$ the active bid and ask stacks for each product. Each existing limit order $i$ is characterized by its price $P_i$ and volume $V_i$. The main decision concerns the traded volumes $v_i$ for order $i$, which are restricted to integer multiples of the minimum bid size $u$. 

Given an initial position $f_h^0$ for contract $h\in \Hc$ and a current fill state of the storage $s_0$, we can now define the intrinsic value as the following MILP:
\begin{align} \label{eq:intrinsic_vanilla}
   & \max \quad && \mathrlap{\sum\limits_{h \in \Hc} \left( \sum\limits_{i \in \Oc_h^-} (P_i - \xi) v_i - \sum\limits_{i \in \Oc_h^+} (P_i+\xi) v_i \right)} && && \text{Revenue} && \\
        & \text{s.t.} \quad	&& 0 \leq v_i \leq V_i, && \forall i \in \Oc_h, \forall h \in \Hc && \text{Order fills} && \notag\\
        & && v_i = k_i u, && \forall i \in \Oc_h, \forall h \in \Hc && \text{Min. bid-size} && \notag\\
        & && f_h^+ = \sum_{i\in \Oc_h^+} v_i, && \forall i \in \Oc_h^+, \forall h \in \Hc && \text{Cum. buy volume} && \notag\\
        & && f_h^- = \sum_{i\in \Oc_h^-} v_i, && \forall i \in \Oc_h^-, \forall h \in \Hc  && \text{Cum. sell volume} && \notag\\
        & && f_{h} = f^0_{h} + f_h^+ - f_h^-, \qquad && \forall h \in \Hc && \text{Final position} && \notag\\
        & && f_h \in [\underline f, \bar f], && \forall h \in \Hc && \text{Power constraints} && \notag\\
        & && f_h = i_h - w_h, && \forall h \in \Hc && \text{Aux. variables} && \notag\\
        & && i_h \in [0, \alpha_h \bar f], && \forall h \in \Hc && \text{Injection constraint} && \notag\\
        & && w_h \in [0, -(1-\alpha_h) \underline f], && \forall h \in \Hc && \text{Withdrawal constraint} && \notag\\
    & && s_{h} = s^- + \eta^+ i_h - ({1}/{\eta^-}) w_h, && \forall h \in \Hc && \text{SOC evolution} && \notag\\
    & && s_{h} \in [0, \bar s], && \forall h \in \Hc && \text{Storage constraints} && \notag \\
        & && \alpha_h \in \{0,1\}, \; k_i \in \Nd , \; \xi \in \mathbb{R}_{\geq 0} && \forall i \in \Oc_h, \forall h \in \Hc, && && \notag
\end{align}
where 
$$ s^- = \begin{cases} s_0, &  \mbox{if $h$ is the first traded contract} \\ s_{h-1}, & \mbox{otherwise.} \end{cases}$$
Note that, to prevent physically impossible simultaneous charging and discharging of the battery during periods of negative prices, we introduce the auxiliary variables $w_h$, $i_h$, and $\alpha_h$. Finally, the variable $\xi$ captures the friction costs associated with higher trading volumes, the corresponding wear on the battery, and trading fees. We express this as $\xi = \xi_{\text{trade}} + \xi_{\text{deg}}$, where $\xi_{\text{trade}}$ denotes the exchange trading fees, and $\xi_{\text{deg}}$ represents a linear approximation of physical battery degradation. Although this degradation model is simplified, it serves as an effective baseline. %By abstracting complex degradation mechanics, we ensure that the core contribution of this paper remains strictly focused on advancing forecasting methodologies and their applications on the continuous intraday market.

The RI policy repeatedly solves the intrinsic problem, either for every market update or after a fixed delay since the last solve has passed. Every solution starts from the initial contractual positions $f_h^0$ from the last solve and a possibly updated initial storage position $s_0$ as input. We refer to \cite{schaurecker2025maximizing} for an exact specification of the RI, including all nuances of bookkeeping for positions, storage levels, and profits.

%%% Section 3.2: The Lazy Clairvoyant
\subsection{The Lazy Clairvoyant}
As mentioned above, the RI greedily sells the battery's flexibility without considering future revenue potential and is, therefore, a simplification of the true value of storage, which would have to be modeled by a dynamic stochastic programming problem.

To correct for this shortcoming, we would need to add an expected value function to the objective of \eqref{eq:intrinsic_vanilla} that models the tradeoff between immediate rewards and future rewards. However, to obtain such a value function, we would have to solve a stochastic optimization problem with a large number of decision stages, high-dimensional randomness, and a large stage space, all of which combined make the problem computationally intractable \citep[see, e.g.,][]{lohndorf2023value}.

Conceptually, the intrinsic's greedy decision-making foregoes part of the optimal value in two ways: Firstly, it does not capture the value of waiting, i.e., the additional profit that can be made by waiting for better prices instead of trading immediately. Secondly, the RI does not capture the value of optimally and repeatedly reshuffling the positions. More specifically, this value arises because the battery operator can trade multiple times rather than just once. In particular, there is the possibility of making recourse decisions based on updated prices at later stages, which must be explicitly considered in an optimal solution to a stochastic programming problem in order to make optimal decisions. In the context of battery trading, this encompasses the possibility of profitably unwinding, extending, or shifting positions taken earlier.
    
In this paper, we focus on the first aspect and use price forecasts to capture part of the value of waiting. To conceptualize our approach, we move away from the stochastic programming formulation of the battery's problem and instead use the clairvoyant's problem as the starting point. A battery operator that is clairvoyant knows future price developments and trades optimally given this knowledge. Since we focus on the value of waiting, we consider what we call a \emph{lazy clairvoyant}, i.e., a fictitious agent who foresees future price developments but is too lazy to reoptimize its decisions, thus not capturing the value of repeated rebalancing of the position. 

A lazy clairvoyant would decide on a storage schedule by considering the best ask and best bid prices in the remaining trading session for each product. In this way, the corresponding policy would plan injection and withdrawal and then trade each product at the optimal time. In a nutshell, our proposed policy is a model predictive control approximation of the lazy clairvoyant. Specifically, we replace the lazy clairvoyant's perfect-foresight prices for each contract with real-time distributional price forecasts.

We illustrate these ideas in the following reduced example, showcasing the decisions of the RI, the lazy clairvoyant, and an adjusted optimization using forecasts. The example has two trading times $t_0$ and $t_1$ at which four hourly products $H_1$, \dots, and $H_4$ can be traded, with prices evolving between the discrete points in time $t_0$ and $t_1$. To keep the problem simple, we assume a storage system with a power capacity of 1 MW and a storage size of 1 MWh, perfect efficiency ($\eta = 1$), and no transaction or degradation costs ($\xi = 0$).

More specifically, in the situation illustrated in Table \ref{tab:example_forecast}, we have the following trading decisions for the four policies.
\begin{table}[htb]
    \centering
    \begin{tabular}{lcccccc}
        \toprule
         & \multicolumn{2}{c}{$t_0$} & \multicolumn{2}{c}{Forecast (for $t_1$)} & \multicolumn{2}{c}{$t_1$} \\
        \cmidrule(lr){2-3} \cmidrule(lr){4-5} \cmidrule(l){6-7}
            & Bid & Ask & Bid & Ask & Bid & Ask \\
        \midrule
        $H_1$ & 28 & 30 & 29 & 31 & 31 & 32 \\
        $H_2$ & 25 & 27 & 28 & 30 & 29 & 32 \\
        $H_3$ & 58 & 60 & 65 & 67 & 66 & 68 \\
        $H_4$ & 63 & 65 & 61 & 63 & 64 & 66 \\
        \bottomrule
    \end{tabular}
    \caption{\label{tab:example_forecast} Best bid/ask quotes and point forecasts for a simplified two-round example. The battery is 1\,MW / 1\,MWh, starts and ends empty, and hours are chronologically ordered $H_1<H_2<H_3<H_4$, so a charge in hour $i$ is only physically feasible ahead of a discharge in hour $j>i$.}
\end{table}

\begin{itemize}
    \item \emph{Rolling Intrinsic:} At $t_0$ the RI takes the largest available spread among feasible pairs: buying (charging) $H_2$ (ask 27) against selling (discharging) $H_4$ (bid 63) yields 36, which beats every other combination. At $t_1$, switching the discharge leg is profitable: unwinding $H_4$ (buy back at 65) and selling $H_3$ instead (bid 66) nets $+1$. Switching the charge leg is not profitable, so the RI keeps its original charge position. The total resulting cash flow is $(-27+63)+(66-65)=36+1=\mathbf{37}$.

    \item \emph{Forecast Augmented RI:} At $t_0$, we compare the current price of each product against its forecast. For charging, $H_2$'s current ask (27) already beats its forecast (30), so we would buy $H_2$ immediately. For the discharging, the current best contract on offer ($H_4$'s bid, 63) is dominated by the forecast for $H_3$ (65), so we leave the position open and wait. At $t_1$ the forecast augmented RI closes the position by selling $H_3$ at the actual realized bid price of 66. Total cash flow is $-27+66=\mathbf{39}$.

    \item \emph{Lazy Clairvoyant:} Without relying on price forecasts, the lazy clairvoyant leverages perfect knowledge of the realized order book across both rounds. For each product, it identifies the best available price across the entire session: the cheapest ask for $H_2$ occurs at $t_0$ (27), while the highest bid for $H_3$ occurs at $t_1$ (66). It buys $H_2$ at $t_0$—carrying an open position, since it does not discharge until $t_1$—and sells $H_3$ once trading resumes. The total cash flow is $66-27=\mathbf{39}$, which is identical to the forecast-augmented RI because the forecast was sufficiently accurate to defer one leg of the transaction.

    \item \emph{Clairvoyant:} In addition to the identical physical schedule, the clairvoyant exploits the fact that it is not otherwise using $H_1$: it can take a purely financial round trip with zero net physical volume by buying at $t_0$'s ask (30) and selling at $t_1$'s bid (31), for $+1$. For $H_4$, a similar arbitrage trade is not possible due to the bid-ask spreads. The total resulting cash flow is $39+1=\mathbf{40}$.
\end{itemize}

This simple example illustrates the potential value of forecasts in rolling intrinsic optimization, which, if sufficiently accurate, can improve trading decisions and increase the cash flow of the storage asset. Furthermore, the example indicates the appropriate target for the forecasts used by the augmented RI: Since the intraday market evolves and reoptimizations happen throughout the trading session, we need forecasts that capture relevant aspects of price evolution over the entire trading session. This implies that average prices widely used in electricity price forecasting, such as ID1 and ID3, are not adequate targets for a useful forecast.
        
Instead, given the mental model of the lazy clairvoyant, the natural targets is the highest bid and the lowest ask over the remaining trading session, at a volume-depth equal to the storage's charge or discharge power. However, as this value is often too volatile for practical use, a more stable target is preferable, as we will discuss in more depth in Section \ref{sec:forecasting}.

%%% Section 3.3: The Augmented Rolling Intrinsic
\subsection{The Augmented Rolling Intrinsic}
As demonstrated in the last section, it might be beneficial to delay trading decisions if prices are likely to move in the storage's favor. In this section, we demonstrate how to use forecasts to adapt the rolling intrinsic policy, to allow the deferral of trades if the forecasts for future prices suggest that this might be more profitable than trading immediately. 

\begin{figure}[h!]
    \centering
    \resizebox{0.4\textwidth}{!}{%
        \definecolor{AskColor}{RGB}{229,84,81}
\definecolor{BidColor}{RGB}{76,118,131}

% Define slightly different colors for the forecast blocks (lighter tint)
\colorlet{forecastAsk}{AskColor!70!white}
\colorlet{forecastBid}{BidColor!70!white}

\tikzset{
  every node/.style={text=white}
}

\begin{tikzpicture}[x=1cm,y=1cm]
  
  % Common dimensions and vertical positions
  \def\boxwidth{2.5}    % cm
  \def\boxheight{0.7}   % cm
  \def\radius{2}        % rounded corner radius
  
  % Vertical positions (adjusted for 4 asks and 4 bids)
  \def\askDot{4.9}  % dotted marker above asks
  \def\askA{4.1}    % 83 \texteuro
  \def\askB{3.3}    % 47 \texteuro
  \def\askC{2.5}    % 39 \texteuro
  \def\askD{1.7}    % 38 \texteuro (Forecast - Lowest Ask)
  
  \def\bidA{0.5}    % 35 \texteuro
  \def\bidB{-0.3}   % 34 \texteuro (Forecast - Interleaved)
  \def\bidC{-1.1}   % 33 \texteuro
  \def\bidD{-1.9}   % 22 \texteuro
  \def\bidDot{-2.7} % dotted marker below bids
  
  % X-offset 
  \def\mainOffset{1.25}  % Center of the main stack

  %%%%%%%%%%%%%%%%%%%%%%%%%%%%%%%%%%%%%%%%%%%%%%%%%%%%%%%%%%%%%%%%%%%%%%%
  % Product h
  \def\x{0}
  \node[text=black] at (\x+\mainOffset, \askDot+0.5) {\large \textbf{Product {\boldmath $h$}}};
  
  % Main Asks:
  \node[text=black] at (\x+\mainOffset, \askDot) {\vdots};
  \node[draw,rounded corners=\radius, fill=AskColor, minimum width=\boxwidth cm, minimum height=\boxheight cm] 
      at (\x+\mainOffset, \askA) {83\,\texteuro, 3 MW};
  \node[draw,rounded corners=\radius, fill=AskColor, minimum width=\boxwidth cm, minimum height=\boxheight cm] 
      at (\x+\mainOffset, \askB) {47\,\texteuro, 11 MW};
  \node[draw,rounded corners=\radius, fill=AskColor, minimum width=\boxwidth cm, minimum height=\boxheight cm] 
      at (\x+\mainOffset, \askC) {39\,\texteuro, 24 MW};
      
  % Forecast Ask (Merged into stack)
  \node[draw, dashed, thick, rounded corners=\radius, fill=forecastAsk, minimum width=\boxwidth cm, minimum height=\boxheight cm] 
      at (\x+\mainOffset, \askD) {38\,\texteuro, $\bar V$ MW};
      
  % Main Bids:
  \node[draw,rounded corners=\radius, fill=BidColor, minimum width=\boxwidth cm, minimum height=\boxheight cm] 
      at (\x+\mainOffset, \bidA) {35\,\texteuro, 17 MW};
      
  % Forecast Bid (Merged into stack)
  \node[draw, dashed, thick, rounded corners=\radius, fill=forecastBid, minimum width=\boxwidth cm, minimum height=\boxheight cm] 
      at (\x+\mainOffset, \bidB) {34\,\texteuro, $\bar V$ MW};
      
  \node[draw,rounded corners=\radius, fill=BidColor, minimum width=\boxwidth cm, minimum height=\boxheight cm] 
      at (\x+\mainOffset, \bidC) {33\,\texteuro, 98 MW};
  \node[draw,rounded corners=\radius, fill=BidColor, minimum width=\boxwidth cm, minimum height=\boxheight cm] 
      at (\x+\mainOffset, \bidD) {22\,\texteuro, 11 MW};
  \node[text=black] at (\x+\mainOffset, \bidDot) {\vdots};

  %%%%%%%%%%%%%%%%%%%%%%%%%%%%%%%%%%%%%%%%%%%%%%%%%%%%%%%%%%%%%%%%%%%%%%%
  % Legend (2x2 Grid enclosed in a box, perfectly centered at the bid-ask spread y=1.1)
  \def\legX{3.8} 
  \def\legY{-0.2} 
  \def\legW{5.6} 
  \def\legH{2.6} 
  
  % Legend Box Background
  \draw[draw=gray!40, thick, fill=gray!5, rounded corners=6pt] (\legX, \legY) rectangle (\legX+\legW, \legY+\legH);
  
  % Column Headers
  \node[text=black, align=center] at (\legX+2.9, \legY+2.1) {\textbf{Ask}};
  \node[text=black, align=center] at (\legX+4.6, \legY+2.1) {\textbf{Bid}};
  
  % Row 1: Market
  \node[text=black, anchor=west] at (\legX+0.3, \legY+1.3) {\textbf{Market}};
  \node[draw, rounded corners=\radius, fill=AskColor, minimum width=1.2cm, minimum height=0.6cm] 
      at (\legX+2.9, \legY+1.3) {};
  \node[draw, rounded corners=\radius, fill=BidColor, minimum width=1.2cm, minimum height=0.6cm] 
      at (\legX+4.6, \legY+1.3) {};
  
  % Row 2: Forecast
  \node[text=black, anchor=west] at (\legX+0.3, \legY+0.5) {\textbf{Forecast}};
  \node[draw, dashed, thick, rounded corners=\radius, fill=forecastAsk, minimum width=1.2cm, minimum height=0.6cm] 
      at (\legX+2.9, \legY+0.5) {};
  \node[draw, dashed, thick, rounded corners=\radius, fill=forecastBid, minimum width=1.2cm, minimum height=0.6cm] 
      at (\legX+4.6, \legY+0.5) {};
  %%%%%%%%%%%%%%%%%%%%%%%%%%%%%%%%%%%%%%%%%%%%%%%%%%%%%%%%%%%%%%%%%%%%%%%

\end{tikzpicture}%
    }
    \caption{Reduced snapshot of the limit order book and forecast orders for product $h$ visible to the optimization.}
    \label{fig:limit-order-book-forecast}
\end{figure}
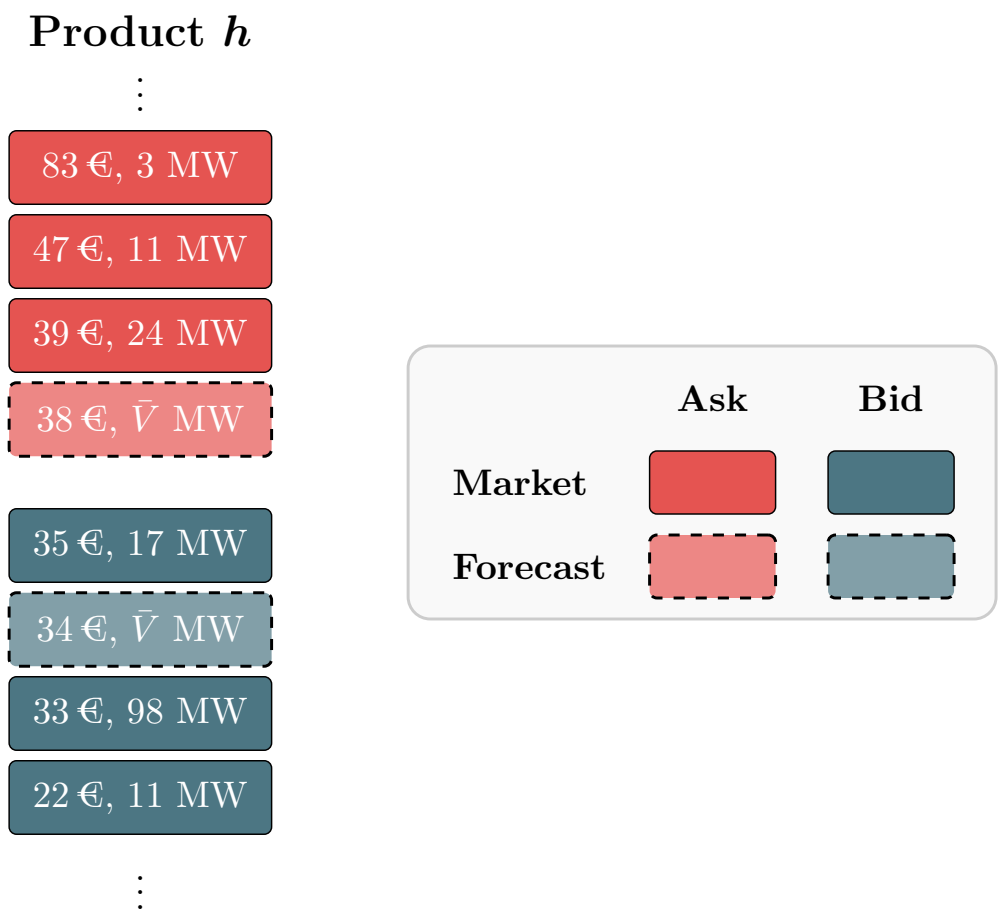

To achieve this without altering the structure of the underlying MILP \eqref{eq:intrinsic_vanilla}, we artificially extend the tradable set of limit orders $\Oc_h^{\pm}$ for each delivery period $h \in \Hc$ with one virtual forecast order $f_h^b$ and $f_h^s$ in the bid and ask stack, respectively. These virtual orders are assigned the forecast buy price $P_{t,h}^{f,+}$ and sell price $P_{t,h}^{f,-}$, with a volume equal to $\bar V = \bar f - \underline f$, as depicted in Figure \ref{fig:limit-order-book-forecast}. 

By feeding the extended order sets $\tilde{\Oc}_h^\pm = \Oc_h^\pm \cup \{f_h^b,f_h^s\}$ directly into \eqref{eq:intrinsic_vanilla}, the optimization automatically evaluates trades against actual orders and virtual trades using virtual forecasting orders, choosing forecast orders if they are more profitable.

When handling the orders with positive $v_i$ from the solution of the optimization problem in backtesting, we place them as limit orders into the order stream after waiting for a certain amount of time, the so-called \emph{technical delay}, which represents delays due to processing in internal systems and submission to the exchange \citep[see][]{schaurecker2025maximizing}.
    
For orders $i\in \Oc_h^\pm$, we additionally use an immediate-or-cancel order qualifier so that they are either immediately cleared against the existing orders currently seen by the intrinsic optimization or discarded. The latter case might occur due to the aforementioned technical delay, which results in changes in the order books relative to the information processed in the last solution instance. Nevertheless, for a reasonably short delay, most of the trades are directly accepted and thus change the battery's contractual delivery schedule and the realized PnL directly.

Quantities in the forecast orders $f_h^b$ and $f_h^s$ represent planned intraday trades for a given delivery period $h$ and cannot be executed immediately against the currently available limit order book. Instead, they end up as resting limit orders at price levels implied by the forecast and volumes dictated by the optimization. 
    
These orders are either accepted by another market participant at some time or changed in subsequent re-optimizations, which may yield different positions in the forecast-based orders as well as different forecast prices. Naturally, the forecast orders influence the PnL and the storage schedule only after they are cleared.

Overall, the discussed augmentation allows the RI to separate trades that should be executed immediately from those that are only attractive at forecast-implied price levels. This approach expresses a view on future intraday prices while still respecting current market liquidity, physical constraints, and, importantly, realistic limit order matching mechanisms that have not yet been discussed for forecasting applications in academic literature. Furthermore, by continuously re-evaluating these orders at millisecond speed, our framework fully enables high-frequency reoptimization at the market's pace while still respecting the information in our continuously updated price forecast.

One important difference between the proposed method and the plain RI is that the latter only optimizes based on existing orders without taking any risk. The decisions of the RI, therefore, always result in non-negative profit and a strictly physically feasible schedule, i.e., a plan of injections and withdrawals that can be executed without the need for further subsequent trading. In contrast, our policy anticipates future price movements and, therefore, sacrifices guaranteed profits and physical feasibility. In particular, if the algorithm chooses to accept existing asks and a forecast bid order or vice versa, it will end up with an imbalanced decision that might even be physically infeasible without further trading. Furthermore, if real price developments deviate from forecasts, it might never be possible to swap the forecast orders with actual orders at a positive PnL, therefore leading to losses for the overall strategy.

To counteract the potential for physical infeasibilities and significant losses due to forecast errors, we introduce two mechanisms that restrict the trading of forecast-based orders. Firstly, we define a gradual dampening of the maximum forecast order size by introducing a linear fading mechanism such that $\bar V$ is replaced by $Q_f := \bar V \phi_h(t)$, where
\begin{equation}\label{eq:fc_fading}
    \phi_h(t) = \max\left(0, \min\left(1, \frac{T_{h,\text{fade}}^{\text{end}} - t}{T_{h,\text{fade}}^{\text{end}} - T_{h,\text{fade}}^{\text{start}}}\right)\right).
\end{equation}
    
The linear weight $\phi_h(t)$ gradually reduces the available volume of the virtual forecast order between $T_{h,\text{fade}}^{\text{start}}$ and $T_{h,\text{fade}}^{\text{end}}$. This prevents sudden, expensive physical schedule corrections near gate closure when an inaccurate forecast causes the rolling intrinsic to leave a position open. \cite{kath2020optimal} motivate such fading mechanisms by demonstrating that linear execution strategies outperform immediate execution in continuous intraday trading.

Furthermore, to mitigate severe losses caused by forecast errors, which can necessitate costly schedule corrections near gate closure, we implement a time-independent stop-loss safeguard: After each execution of the forecast-augmented optimization, its objective value is evaluated against a counterfactual, market-only baseline that excludes virtual forecast orders ($f_h^b$ and $f_h^s$). If the forecast-driven objective exceeds the market-only reward by more than a user-defined threshold $\gamma$ (in EUR), the stop-loss triggers. In this case, the forecast-based solution is discarded in favor of the counterfactual market-only orders, and virtual forecast orders are permanently deactivated for all active contracts for the remainder of their trading sessions. This safeguard is designed to limit the impact of rare but severe forecast failures, particularly during extreme price episodes when incorrect forecast-driven trades can cause very expensive corrections. We detail the calibration of this threshold in Section~\ref{sec:simulation-results-setting}.

%%%%%%%%%%%%%%%%%%%%%%%%%%%%%%%%%%%%%%%%%%%%%%%%%%%%%%%%%%%%%%%%%%%%%%
%%% Section 4: Forecasting the Empirical Distribution of Midprices %%%
%%%%%%%%%%%%%%%%%%%%%%%%%%%%%%%%%%%%%%%%%%%%%%%%%%%%%%%%%%%%%%%%%%%%%%
\section{Forecasting the Empirical Distribution of Midprices}\label{sec:forecasting}
The augmented rolling intrinsic employs a forecast as a proxy for the achievable price in the market. Therefore, the natural next step is the development of an appropriate target variable. As argued above, probabilistic forecasting of volume-weighted average prices is not optimal for battery storage optimization. This aggregation reduces the information since decision-making is taking place at the order level --- the atomic unit of the market. 

However, modeling the distribution of a volume-weighted average price is different from modeling the distribution of all prices, where the latter is of interest for decision-making. An attractive alternative that avoids averaging has been proposed by \cite {janke2019forecasting} by forecasting the volume-weighted empirical cumulative density function (ECDF) of the trades in the market. A similar approach has been taken by \cite{berrisch2023high} to forecast the minimum and maximum load in a distribution grid during a time interval. This approach is naturally aligned with the setting of the lazy clairvoyant, as we desire a forecast that provides an estimate of the buy/sell price that can be achieved in the remaining trading session. 

In this paper, we forecast the ECDF of mid prices and discuss the conceptual advantages of this approach, its connection to the Wasserstein distance and the practical implementation in the following Sections~\ref{sec:forecasting-target-variable} and~\ref{sec:forecast-wasserstein} before introducing the models and data formally in Sections~\ref{sec:limit-order-book-features} to~\ref{sec:data-intraday-renwables}.

%%% Section 4.1: The Distribution of Mid Prices as Target Variable
\subsection{The Distribution of Mid Prices as Target Variable}\label{sec:forecasting-target-variable}
To approach the question of an appropriate target variable, let us assume that we have perfect information about all intraday prices and return to the \emph{myopic}, the \emph{clairvoyant} and the \emph{lazy clairvoyant} model of Section~\ref{sec:introduction} and \ref{sec:optimization-with-forecasts}. The myopic optimization only has information about the current market price. The clairvoyant optimization has perfect information about all future prices. The lazy clairvoyant also has all the information, but due to its laziness, does not rebalance and therefore only uses information about the minimum and maximum prices in the remaining trading session. 

Intuitively, one might conclude that forecasting the minimum and maximum prices of the trading session is sufficient since the storage operator would only buy at the minimum price and sell at the maximum price. However, these extreme prices are typically hard to predict. We can alleviate this drawback by considering the empirical quantiles~$q_l = \alpha / 2$ and~$q_u = 1 - \alpha / 2$ of the midprices instead.  Choosing an appropriate quantile level for buying and selling introduces a new hyperparameter that controls the balance between taking few large profit opportunities and many small trades. Note that the quantile set can, but does not need to be symmetric; however, non-symmetric intervals increase the cost of tuning quadratically. 

\begin{figure}[t]
    \centering
    \includegraphics[width=\linewidth]{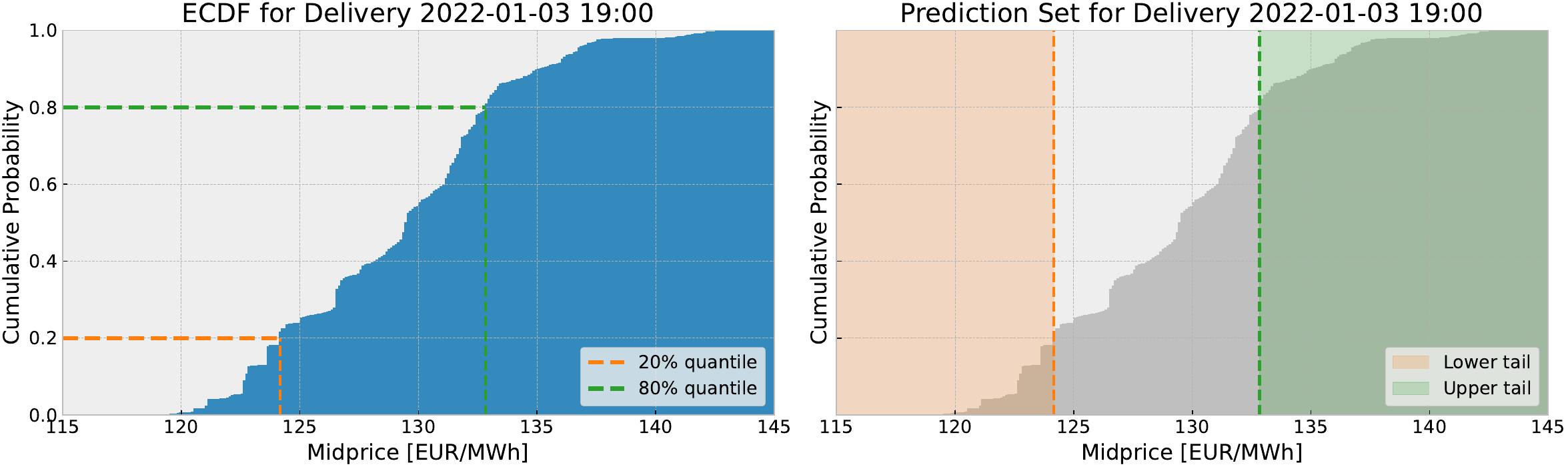}
    \caption{
        Definition of the buying and selling prices based on the empirical distribution of all mid prices. 
        For the exemplary $\alpha = 0.4$ depicted here, we receive limit prices for selling and buying $(q_l, q_u)$ that will lead to a trade if we approach prices in the tail mass of the distribution.
    }
    \label{fig:ecdf-prediction-set}
\end{figure}

Figure \ref{fig:ecdf-prediction-set} illustrates that we are interested in the realized observations outside of $({F}^{-1}(q_l), {F}^{-1}(q_u))$, where ${F}$ is the cumulative distribution function of the midprices until gate closure. The idea is that the rolling intrinsic places, for example, a bid at a price for which it expects that at least $({\alpha}/{2})$-percent of the prices in the remaining trading session will be lower.

When replacing perfect information with forecasts, we anticipate that the predicted quantiles might not be achieved in the market, especially for small $\alpha$. Therefore, $\alpha$ also plays a role in controlling the risk of the forecast, and its choice must be calibrated to the operator's risk appetite and the quality of the forecasts. Like every other hyperparameter, $\alpha$ can be optimized on a validation set. Note that in probabilistic forecasting for a target~$y$, prediction intervals are commonly defined as the area containing~$y_i$ with probability~$1 - \alpha$, capturing aleatoric uncertainty about a single observation~$y_i$ \citep{hullermeier2021aleatoric}. This differs from the distributional setting in which we operate in this paper, where we are interested in quantifying the shape of the distribution of~$y$.

\begin{figure}[t]
    \centering
    \includegraphics[width=\textwidth]{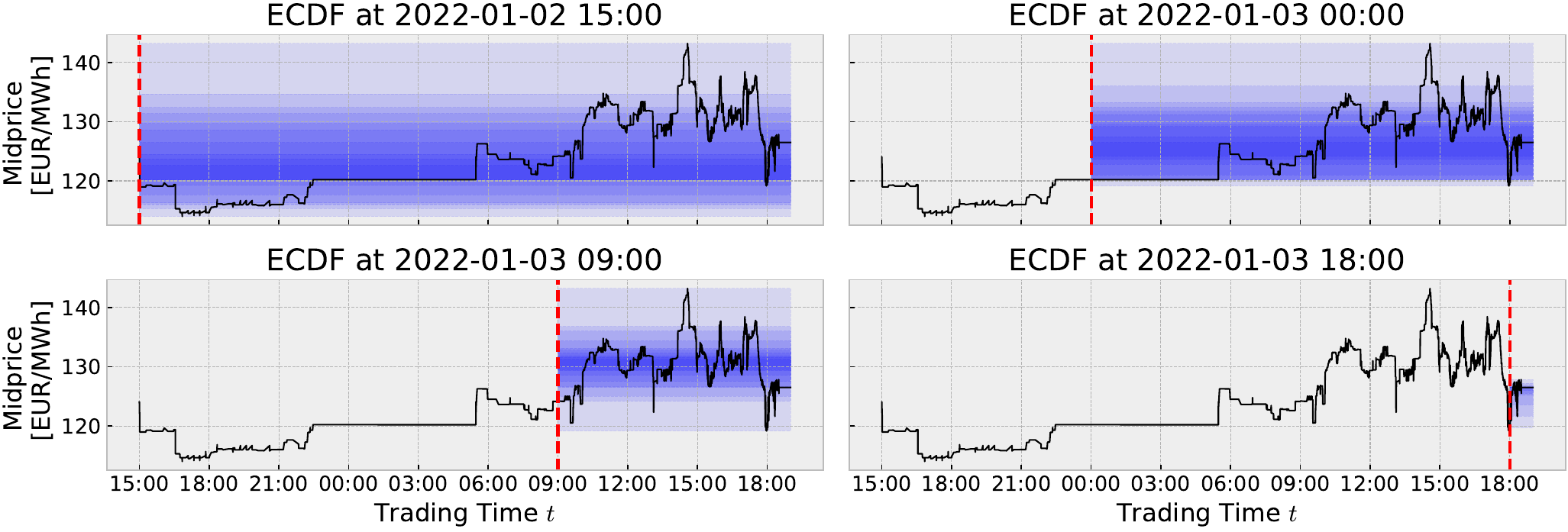} \\ [1em]
    \includegraphics[width=\textwidth]{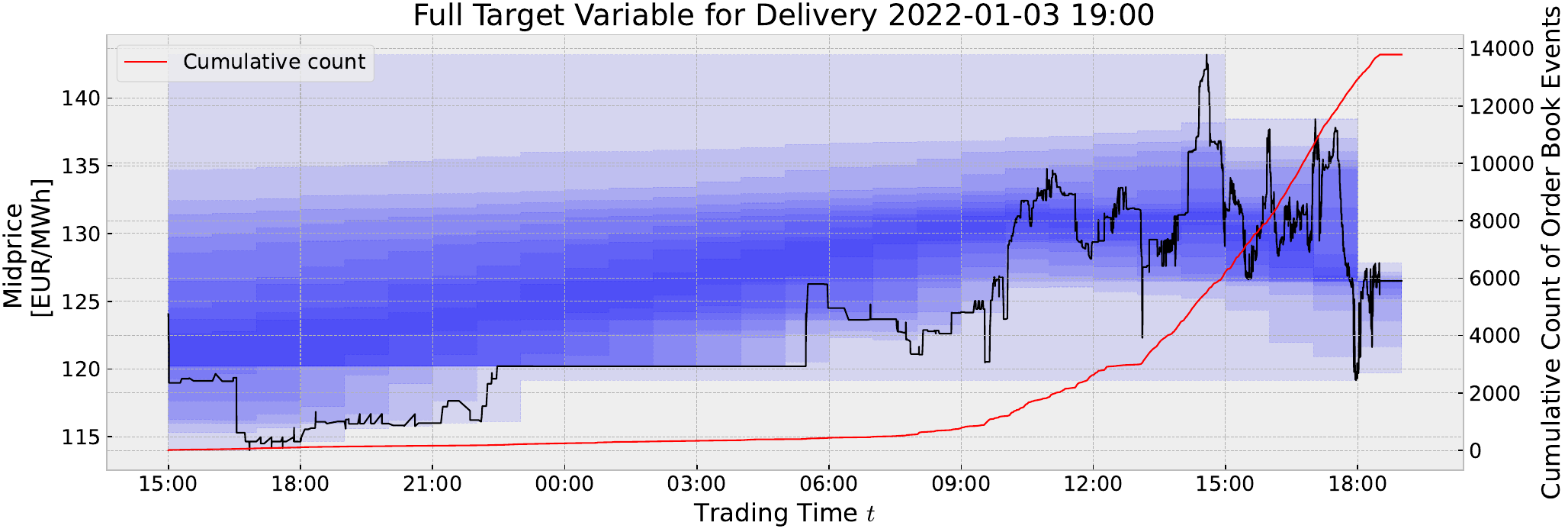}
    \caption{
        Construction of the target variable. The top four panels show the ECDF of all mid-prices after a certain trading time $t$. The larger bottom panel gives the full target variable constructed by hourly slices of ECDFs.
    }\label{fig:ecdf-by-period}
\end{figure}

While the trading session progresses, the forecast augmented RI continues to make decisions based on the available information about future prices. As time progresses, the set of future prices yet to be observed decreases as market prices are realized. This leads to a need to update the (predicted) distribution of prices. Therefore, we update the distribution regularly, taking into account all midprices in the \emph{remaining} trading session of the respective delivery hour~$h$ as is depicted in Figure~\ref{fig:ecdf-by-period}. The full information given to the optimization, therefore, contains the distribution of prices in the remaining session at each (discrete) point of forecasting. This gives the target variable for forecasting shown in the lower panel of Figure~\ref{fig:ecdf-by-period} and further emphasizes the difference between a classical probabilistic forecasting setting, in which the target variable is a scalar value, and the distributional setting of this paper.

In the context of intraday order book data, forecasting the empirical distribution of mid prices has the additional advantage that it allows for efficient information compression of the high-volume of data. Instead of forecasting every single order book update or every other trade, we have a compact representation of the price distribution during the trading session that retains the information necessary for efficient decision-making.

%%% Section 4.2: Connecting Learning the ECDF to the Wasserstein Distance
\subsection{Connecting Learning the ECDF to the Wasserstein Distance}\label{sec:forecast-wasserstein}
In the following, we briefly argue that a classic least absolute deviation respectively least-squares on the empirical quantile minimizes the Wasserstein distance (WD) between the true ECDF and the predicted ECDF. \cite{petersen2019frechet} discuss the theoretical properties of the regression of vector-valued predictors and distributional targets from a theoretical perspective. We provide a brief, intuitive argument here, aligned with the works of  \cite{janke2019forecasting} and \cite{berrisch2023high}. 

The Wasserstein Distance on the line $\operatorname{WD}_p(F_0, F_1)$ between two distributions $F_1$ and $F_2$ is given by \begin{equation}
    \operatorname{WD}_p(F_1, F_2) = \left(\int_0^1 \mid F_1^{-1}(q) - F_2^{-1}(q) \mid^p \mathrm{d}q \right)^{1/p}
\end{equation}
which can be approximated by a dense grid of quantiles $\mathcal{Q}$ \begin{equation}
     \operatorname{WD}_p(F_1, F_2) \approx \left(\frac{1}{|\mathcal{Q}|}\sum_{q \in \mathcal{Q}} \mid F_1^{-1}(q) - F_2^{-1}(q) \mid^p \mathrm{d}q  \right)^{1/p}
\end{equation}
Denote the true EDCF of the target variable as ${F}(q)$ and the predicted ECDF as $\widehat{{F}}(q)$.
We use a machine learning method $h(\mat{X}) = \hat{\mat{y}}$ to minimize the $\ell_2$-loss between the prediction $\hat{\mat{y}}$ and the realized values $\mat{y}$, \begin{equation}
    \min_{h \in \mathcal{H}} \| h(\mat{X}) - \mat{y} \|_2^2.
\end{equation}
Inserting the $q$ quantile of the ECDF $h_q$ of as target implies that we estimate the model $h^*_q(\mat{X})$  \begin{equation}
    \min_{h_q \in \mathcal{H}} \| h_q(\mat{X}) - {F}^{-1}(q) \|_2^2.
\end{equation}
and hence 
\begin{equation}
    \operatorname{WD}_2({F}, \widehat{{F}}) \approx \left(\frac{1}{|\mathcal{Q}|}\sum_{q \in \mathcal{Q}} \mid {F}^{-1}(q) - h^*_q(\mat{X}) \mid^2  \right)^{1/2}
\end{equation}
is minimized by the models $h^*_q(\mat{X})$ for all $q \in \mathcal{Q}$. The same argument holds for the $\ell_1$-loss and $\operatorname{WD}_1$. 

By this argument, we can see that minimizing the least absolute deviation or least-squares loss on the empirical quantiles is equivalent to minimizing the Wasserstein distance between the true and predicted ECDF. A clear advantage of learning the ECDF is that we have access to a large statistical and machine learning toolbox for learning $h(\mat{X})$. The trade-off for this flexibility is the issue that quantiles are not necessarily non-decreasing, since we have separate models for the levels $q$, which can lead to inconsistencies in the predicted ECDF. We address the issue by sorting, similar to common approach in quantile regression to ensure non-crossing quantiles \citep[see e.g.][]{chernozhukov2010quantile, park2022learning}. Alternatively, one could also employ more sophisticated methods to ensure non-crossing quantiles through learning with monotonicity constraints \citep{park2022learning, yu2025orderfusion}, which we leave for future research.

%%% Section 4.3: Limit Order Book Data and Features
\subsection{Limit Order Book Data and Features}\label{sec:limit-order-book-features}
\begin{figure}[thb]
    \centering
    \includegraphics[width=0.8\textwidth]{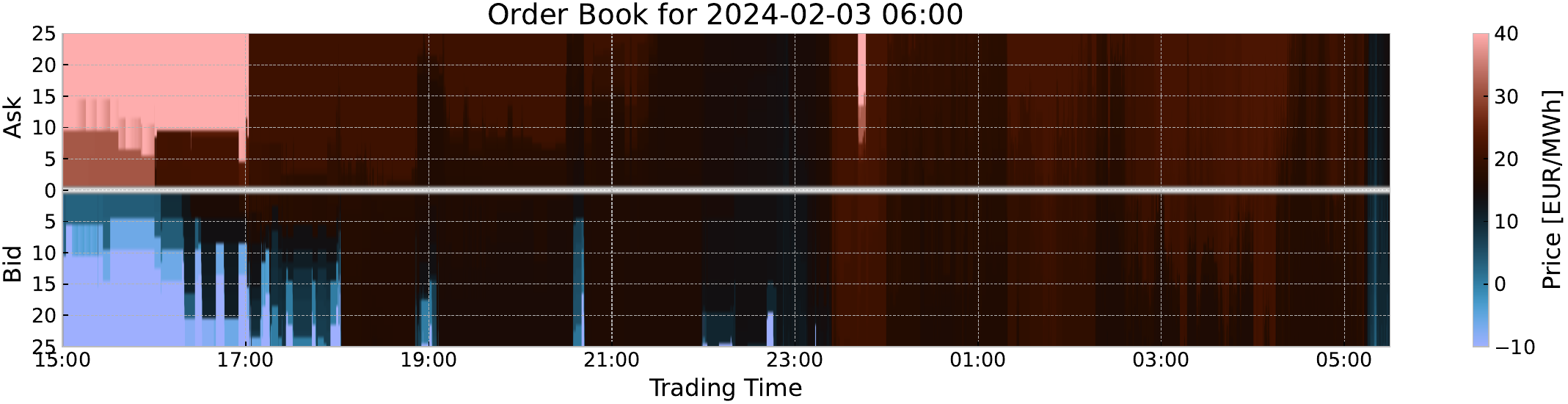}
    \includegraphics[width=0.8\textwidth]{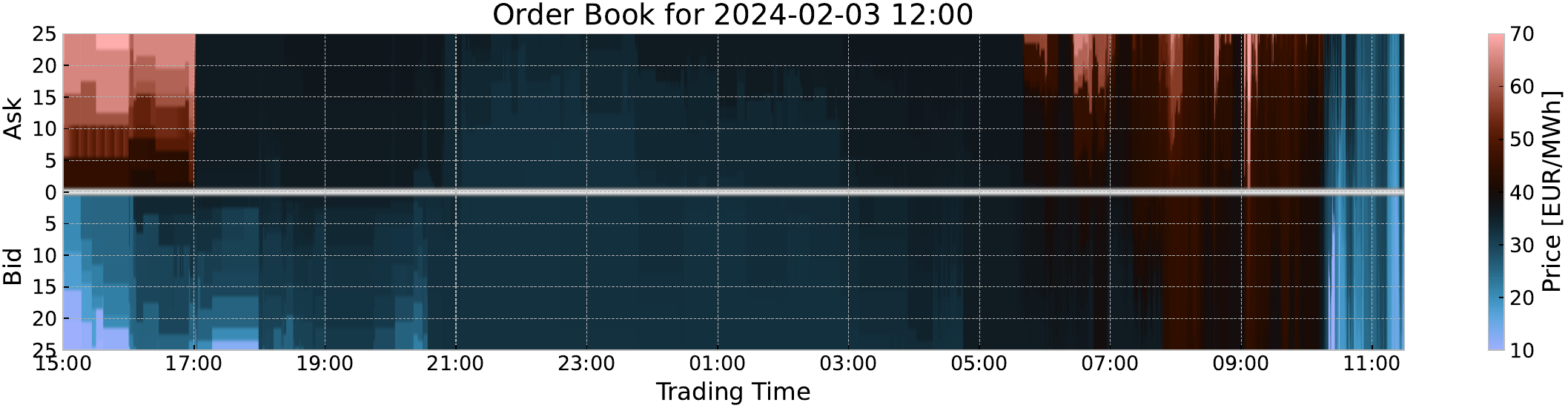}
    \includegraphics[width=0.8\textwidth]{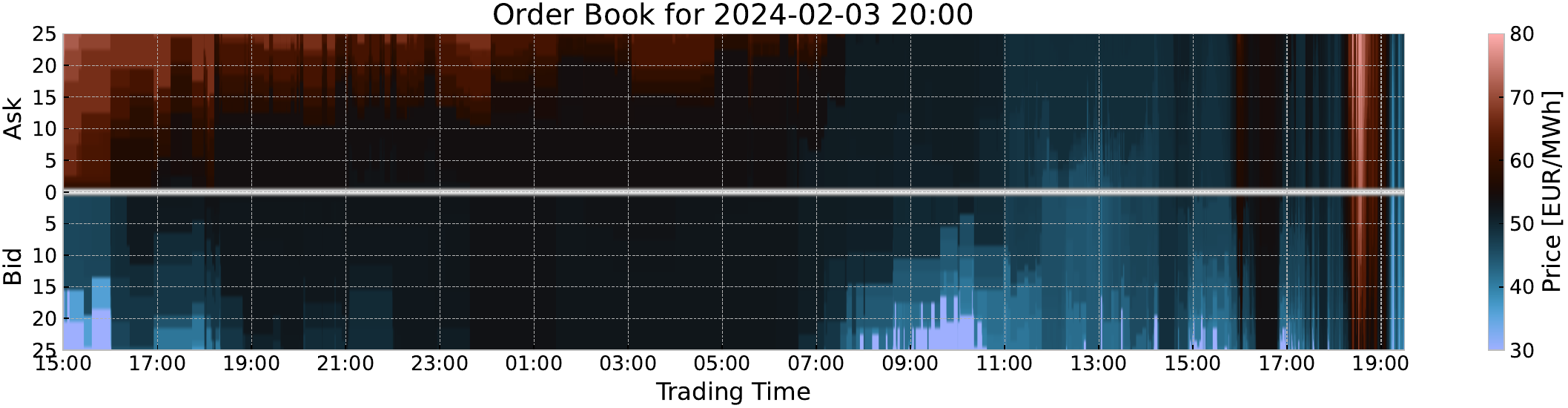}
    \caption{Visualization of the order book for delivery at 06:00, 12:00, and 20:00 on 3rd of February, 2024. The color denotes the price, the bid/ask volumes are given on the $y$-axis. Note that the colorscale changes throughout the panels.}
    \label{fig:order-book-example}
\end{figure}

We use a dataset of all limit order book events in the German intraday electricity market from 2022 to 2024 handled through the trading platform of EPEX SPOT SE. 
We preprocess the data by recreating the limit order book for each delivery period $h$ at each event time $t$, as shown in Figure \ref{fig:order-book-example}. 
Subsequently, we focus on the mid price of the limit order book. This is an improvement over previous research on battery optimization, which only employs realized transaction data \citep{miskiw2025coordinated, oeltz2025rolling}.

Since $t$ might be unevenly spaced, we aggregate the data on an evenly spaced grid $\mathcal{G}$ by building a time-weighted empirical cumulative distribution function and time-weighted average mid prices \citep{grindel2022forecasting}. The grid has an hourly frequency between the market opening $\bar{g}$ and the start of physical delivery $h$, hence $\mathcal{G} = \{\bar{g}, g_1, ..., g_m, h\}$ with a distance of $\pi = 60$ minutes. We denote the quoted price for the $v$-th MW as $P^{[v,s]}_{h,t}$, where $s \in \{\text{bid}, \text{ask}\}$. The time-weighted ECDF on the grid $\mathcal{G}$ is calculated as \begin{align}
    \underbrace{
        \bar{Q}^{q}_{h,g} = F^{-1}_q\left(\left\{\dfrac{P^{[1,\text{bid}]}_{h,t} + P^{[1,\text{ask}]}_{h,t}}{2}\right\} \forall t \in [g-\pi, g]\right)
    }_\text{Time-weighted ECDF of the previous grid interval.} 
    && \text{and} && 
    \underbrace{
        Q^{q}_{h,g} = F^{-1}_q\left(\left\{\dfrac{P^{[1,\text{bid}]}_{h,t} + P^{[1,\text{ask}]}_{h,t}}{2}\right\} \forall t \in [g, h]\right)
    }_\text{Time-weighted ECDF of the remaining trading session.}
\end{align}
where the quantile operator $F^{-1}_q$ is weighted by the inter-arrival time~$\Delta{}t = t_2 - t_1$ for two consecutive events. That is, we take the quantiles of all observed mid prices, weighted by the time between the observations proportional to probabilities. In the limit order book, this represents the time an order was on the screen and therefore available for market participants to execute.

Additionally, we generate features from the limit order book data. The time-weighted-average price for delivery hour $h$, side $s \in \{\text{bid}, \text{ask}\}$ and volume $v$ at grid point $g$ is defined as 
\begin{equation}\label{eq:weighted_price}
    \bar{P}^{[v,s,\pi]}_{h,g} = \frac{1}{\pi} \sum_{t \in [g-\pi, g]} \Delta{}t P^{[v,s]}_{h,t}
\end{equation}
where $P^{[v,s]}_{h,t}$ is the price quoted at time $t$ for delivery hour $h$ and $\pi$ is the length of the time window. Based on this, we can calculate the time-weighted order book statistics according to Table \ref{tab:order-book-features}. Similar feature constructions have been proposed by \cite{grindel2022forecasting} and, in the context of trade data, by \cite{shinde2021analysing,yu2025orderfusion}.

%%% Section 4.4: Forecasting Models
\subsection{Forecasting Models}\label{sec:forecasting-models}
We employ a set of six different forecast models to predict the quantiles~$\widehat{Q}^{q}_{h,g}$ of the empirical distribution of mid prices~$F$. Based on the day-ahead and intraday price and the ECDF of all mid prices in the trading session up to the forecasting time, we develop three simple benchmark models that add historically observed residuals to the respective point forecasts to create models for the ECDF (see Appendix~\ref{app:benchmarks} for details). Additionally, we develop two regression models and the perfect foresight model as a benchmark. This allows us to evaluate the quality of the forecast approximation of the \emph{lazy clairvoyant} by the augmented rolling intrinsic.

We denote the perfect foresight forecast as \textsf{PerfectForesight}, the benchmarks as the \textsf{NaiveDayAhead}, \textsf{NaiveIntraday} and \textsf{NaiveECDF} in the following Sections~\ref{sec:forecasting-results} and~\ref{sec:simulation-results}. We focus in the regression models in the following. 

Our first basic regression model is defined as
\begin{equation}
    \widehat{Q}^q_{h,g+1} 
        = \beta_0 
        + \beta_1 P^\text{DA}_h 
        + \beta_2 \bar{P}^{[\text{mid}, \pi]}_{h,t} 
        + \beta_3 \widehat{\Delta\operatorname{RES}}_{h,t}
        + \beta_4 \operatorname{ResLoad}_h
        + \beta_5 Q^q_{h,g},
\end{equation}
while our extended regression model is of the form
\begin{align}
    \widehat{Q}^q_{h,g+1} 
        &= \beta_0 
        + \beta_1 P^{DA}_h
        + \sum_{i=-2}^2\beta_{4+i} \bar{P}^{[\text{mid}, \pi]}_{h+i,t} 
        + \beta_7 \text{BAS}^{[\pi]}_{h,t}
        + \sum_{a \in \mathcal{Q}} \beta_{8+a}\widetilde{Q}^a_{h,t} 
        + \sum_{i=-2}^2\beta_{11+i} \text{M}^{[\text{DA}, \pi]}_{h+i,t} 
        + \sum_{i=-2}^2\beta_{15+i} \text{M}^{[\text{ID}, \pi]}_{h+i,t} \\ \nonumber
        &+ \sum_{s \in \text{bid}, \text{ask}} \sum_{v \in 5, 10, 25, 50} \beta_{17+i} \text{S}^{[v, s, \pi]}_{h,g}
        + \beta_{18} \widehat{\Delta\operatorname{RES}}_{h,g}
        + \beta_{19} \operatorname{ResLoad}_h 
        + \sum_{i=1}^7 \beta_{20+i} \operatorname{DOW}(d) \\ \nonumber
        &+ \beta_{27} \operatorname{Gas}_{d(h)}
        + \beta_{28} \operatorname{Coal}_{d(h)}
        + \beta_{29} \operatorname{EUA}_{d(h)}
        + \beta_{30} \operatorname{Oil}_{d(h)},
\end{align}
which is estimated using the least absolute shrinkage and selection operator (LASSO) to perform variable selection and regularization. $\pi$ always denotes a window size, see Equation \ref{eq:weighted_price}. All distributional models are estimated for all $q \in \mathcal{Q}$, where $\mathcal{Q} = \{0.05, ..., 0.95\}$.

Table \ref{tab:order-book-features} provides an overview of all features in the model and their respective calculations. The model consists of common energy market features such as fundamental forecasts and fuel prices, as well as market-specific features such as prices and order book statistics \citep{yu2025orderbook, grindel2022forecasting}. We choose the hyperparameter for the LASSO by selecting the model with the lowest Bayesian information criterion (BIC) on the training set. The models are denoted as \textsf{OLS} and \textsf{LASSO} in the remainder of the text. 
We omit the subscripts $h, t$ for $\beta_i$ for the sake of readability. 

We employ a local approach to forecasting, as is common in electricity price forecasting \citep{grindel2022forecasting,puc2024corrected,marcjasz2023trading}, which implies that we estimate a separate model for each quantile $q \in \{0.05, 0.1, ..., 0.95\}$ and for each issue time $g$.  
%Accounting for the fact that we issue the first forecast after the first hour of trading and do not need a forecast for the last hour of trading, we issue $24 \times 7 + (24 \times 23) / 2 = 444$ forecasts of the future ECDF of all mid prices and therefore utilize $444 \times 19$ models. Note that the products that deliver at the beginning of the day trade shorter than those that trade at the end of the day. 
To allow for daily refitting of all models, we employ the online learning algorithm for the LASSO by \cite{angelosante2009online,messner2019online, hirsch2024online} implemented in the \texttt{ondil} Python package.\footnote{See: https://github.com/simon-hirsch/ondil.} Table \ref{tab:order-book-features} gives an overview of all features in the model and their respective calculations.

\begin{table}[t]
    \centering
    \resizebox{\textwidth}{!}{%
        \begin{tabular}{rll}
            \toprule
            Variable            & Calculation & Brief Description \\
            \midrule
            Bid-ask spread      
                &   $\text{BAS}^{[\pi]}_{h,g} = \bar{P}^{[1,\text{ask}, \pi]}_{h,g} - \bar{P}^{[1,\text{bid}, \pi]}_{h,g}$ 
                & The time-weighted bid-ask spread at 1 MW depth of the last $\pi = 5$ minutes. \\ [1ex]
            Mid-Price
                &   $\bar{P}^{[\text{mid}, \pi]}_{h,g} = ( \bar{P}^{[1,\text{ask}, \pi]}_{h,g} + \bar{P}^{[1,\text{bid}, \pi]}_{h,g})/2$
                & The time-weighted mid-price of the last  $\pi = 5$ minutes. \\ [1ex]
            Order book steepness    
                &   ${\text{S}}^{[v, s, \pi]}_{h,g} = \bar{P}^{[v,s, \pi]}_{h,g} - \bar{P}^{[1,s, \pi]}_{h,g}$ 
                & The order book steepness of the last $\pi=5$ minutes. \\ [1ex] % maybe this should be absolute? 
            Momentum            
                &   $\text{M}^{[\text{ID},\pi]}_{h,g} = \bar{P}^{[\text{mid}, \pi]}_{h,g-1+\bar{\pi}} -\bar{P}^{[\text{mid}, \pi]}_{h, g-\bar{\pi}}$ 
                &   Price changes in the last $\pi = 5$ minutes. \\ [1ex]
               
                &   $\text{M}^{[\text{DA}, \pi]}_{h,g} = P^{\text{DA}}_{h} - \bar{P}^{[\text{mid}, \pi]}_{h, g-\bar{\pi}}$ 
                & Price changes since the day-ahead with $\pi=5$ minutes. \\ [1ex]
            Residual Load       &   $\widehat{\text{ResLoad}}_{h,g}$ & Difference between forecasted total system load and renewable production. \\ [1ex]
            Forecast changes.    &   $\widehat{\Delta\operatorname{RES}}_{h,g}$ & Change in renewable energy production forecasts since the day-ahead. \\ [1ex]
            Fuel Prices.         &   $\operatorname{Gas}_d, \operatorname{Coal}_d, \operatorname{EUA}_d, \operatorname{Oil}_d$ & Day-ahead fundamental fuel prices (TTF for gas, API2 for coal, Brent Oil). \\
            \bottomrule
        \end{tabular}%
    }%
    \caption{Fundamental features and Order book statistics calculated on the time-weighted average prices. $\pi$ denotes the length of the time window for the time-weighted average price (in minutes). For the calculation of the momentum, we use the time-weighted average prices for the first and last 5 minutes within one hour, which gives a stable representation of price changes between since the last forecast on the hourly grid $\mathcal{G}$.}\label{tab:order-book-features}
\end{table}

%%% Section 4.5: Intraday Renewable Energy Production Forecasts
\subsection{Intraday Renewable Energy Production Forecasts} \label{sec:data-intraday-renwables}
Commonly, market participants in the short-term electricity market have access to continuously updated forecasts for renewable energy production. The forecasts used in this paper are produced by an international energy meteorology company~(which chooses to remain anonymous) and are provided by Statkraft Trading GmbH. The data set contains renewable energy production forecasts for onshore and offshore wind as well as solar assets in Germany. The forecasts have an hourly resolution, are denoted in MWh, and are updated every hour. The data set spans 2022 to 2024. For two missing days, we use ENTSO-E day-ahead forecasts (published before $d-1$, 12:00 hours) and intraday forecasts (published $d$, 10:00 hours). Single missing values are forward-filled (which is the same as assuming that no new data arrived in the meantime). Overall, there are fewer than 800 missing values for approximately 360,000 observations before preprocessing the data (0.2\% missing data).

\begin{figure}[htb]
    \centering
    \includegraphics[width=0.8\textwidth]{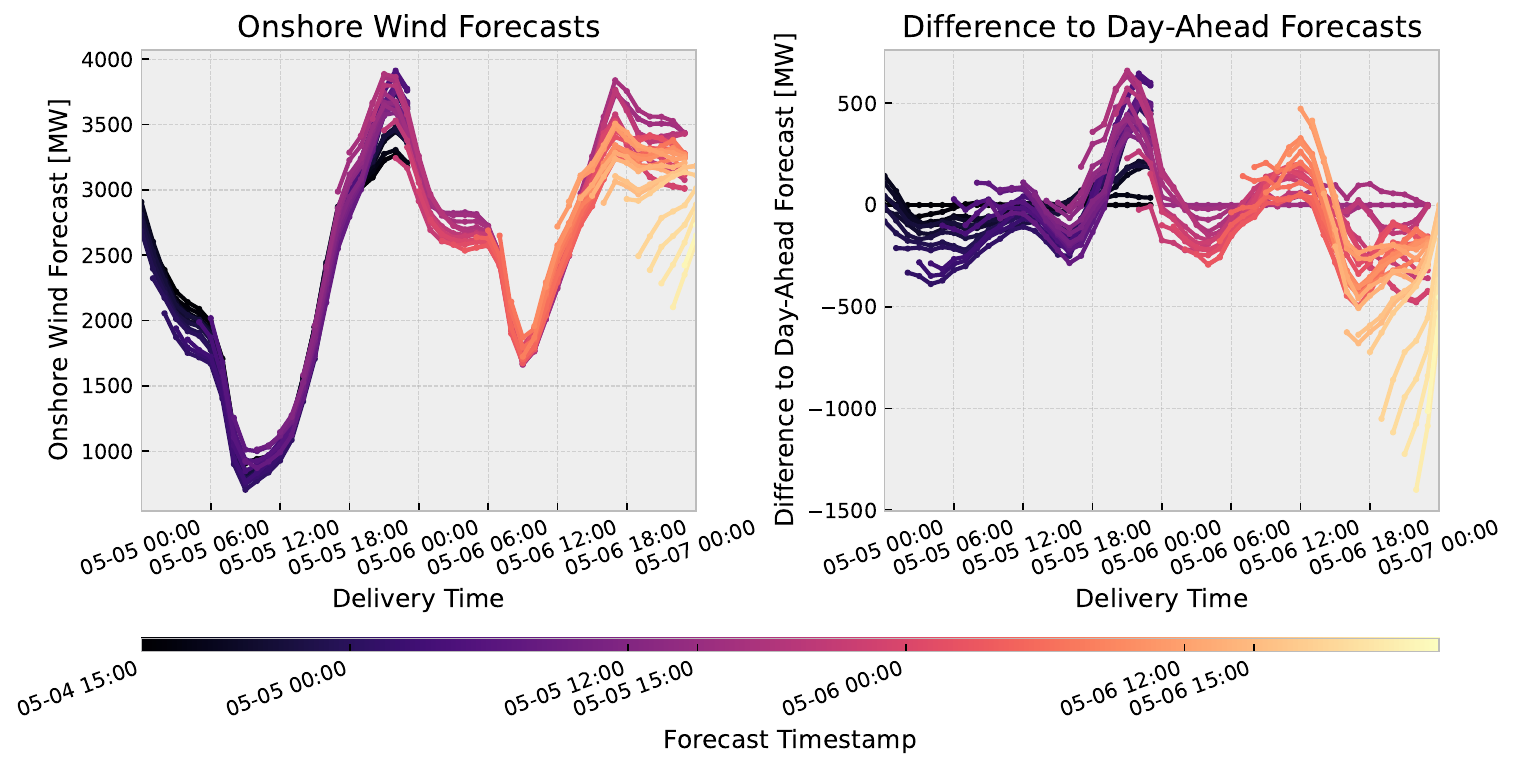}
    \caption{Example of renewable energy production forecasts for wind onshore, offshore and solar assets in Germany. The forecasts are denoted in MWh and updated every hour.}
    \label{fig:forecast_data}
\end{figure}

Figure \ref{fig:forecast_data} shows exemplary forecasts for the wind offshore production for 5th and 6th of May 2024. Since the intraday market is driven by the change in fundamentals since the day-ahead \citep{ziel2017modeling, kulakov2021impact, marcjasz2023trading,serafin2022trading}, we process forecasts by subtracting the day-ahead forecast from the most recent forecast. For each delivery hour $h$ and grid point $g$ we select the last forecast issued before $\max(t) \leq g$ and denote it as \begin{equation}
    \widehat{\Delta\text{WindOn}}_{h,g} = \widehat{\text{WindOn}}_{h,g} - \widehat{\text{WindOn}}_{h,\bar{g}}
\end{equation}
where $\bar{g}$ denotes the day-ahead forecast, and similarly for offshore wind $\widehat{\Delta\text{WindOff}}_{h,g}$ and solar production $\widehat{\Delta\text{Solar}}_{h,g}$. We summarize the forecast changes for all technologies 
$$\widehat{\Delta\operatorname{RES}}_{h,g} = \widehat{\Delta\text{WindOff}}_{h,g} + \widehat{\Delta\text{WindOn}}_{h,g} + \widehat{\Delta\text{Solar}}_{h,g},$$ which is the change in renewable energy production forecasts since the day-ahead. The residual load is defined as 
$$
    \widehat{\text{ResLoad}}_{h,g} = \widehat{\text{Load}}_h - \widehat{\text{Res}}_{h,g}, 
$$
that is the difference between the total system load and the renewable production. The system load forecast of ENTSO-E is only available at the day-ahead stage.

%%%%%%%%%%%%%%%%%%%%%%%%%%%%%%%%%%%%%%
%%% Section 5: Forecasting Results %%%
%%%%%%%%%%%%%%%%%%%%%%%%%%%%%%%%%%%%%%
\section{Forecasting Results}\label{sec:forecasting-results}
This section describes the results from a statistical evaluation perspective. We discuss metrics in the following Section \ref{sec:forecasting-results-metrics} and present results in Section \ref{sec:forecasting-results-results}. We discuss the implications of the forecasting results for optimization in Section~6.

%%% Section 5.1: Evaluation Metrics
\subsection{Evaluation Metrics}\label{sec:forecasting-results-metrics}
We consider the $L_1$ and $L_2$ errors of the predicted quantiles as error metrics, which are equivalent to the $\operatorname{WD}_1$ and $\operatorname{WD}_2$ between the true and predicted ECDF. Additionally, we evaluate the average width of the prediction intervals. 
\cite{genov2025balancing} and \cite{godahewa2025forecast} show that changing the forecast often introduces switching costs; hence, forecast stability is an important aspect to consider in the context of optimization. We evaluate the stability of the forecasts by calculating the average directional change and the average absolute change of forecasts. For a forecast for delivery period $h$ issued at time $g$ for quantile level $q$, we denote the forecast as $\widehat{Q}^q_{h,g}$ and the realized value as $Q^q_{h,g}$. 
% In summary, the standard error metrics are given by:
% \begin{align}
%     \text{MAE}^q_{h,g} = |\widehat{Q}^q_{h,g} - Q^q_{h,g}|; &&  
%     \text{RMSE}^q_{h,g} = (\widehat{Q}^q_{h,g} - Q^q_{h,g})^2; &&
%     \text{Width}^q_{h,g} = \widehat{Q}^{q_u}_{h,g} - \widehat{Q}^{q_l}_{h,g};  && 
%     \text{Bias}^q_{h,g} = \widehat{Q}^q_{h,g} - Q^q_{h,g}.
% \end{align}
% Furthermore, we evaluate 
% \begin{align}
%     \text{RatioOverprediction}^q_{h,g} &= \frac{1}{T}\sum_{t=1}^T\mathbb{I}(\widehat{Q}^q_{h,g} - Q^q_{h,g} > 0) \\
%     \text{Direction}^q_{h,g} &= \frac{1}{T}\sum_{t=1}^T\operatorname{sign}(\widehat{Q}^q_{h,g} - \widehat{Q}^q_{h,g-1}) \\
%     \text{Stability}^q_{h,g} &= |\widehat{Q}^q_{h,g} - \widehat{Q}^q_{h,g-1}|.\label{eq:stability_1}
% \end{align}
The standard error metrics are given by:
\begin{align}
    \text{AE}^q_{h,g} = |\widehat{Q}^q_{h,g} - Q^q_{h,g}|; &&  
    \text{SE}^q_{h,g} = (\widehat{Q}^q_{h,g} - Q^q_{h,g})^2; &&
    \text{Width}^\alpha_{h,g} = \widehat{Q}^{1-\alpha/2}_{h,g} - \widehat{Q}^{\alpha/2}_{h,g};  && 
    \text{Bias}^q_{h,g} = \widehat{Q}^q_{h,g} - Q^q_{h,g}
\end{align}
and aggregated by averaging (and applying the square root) over the indices~$g, q$ or~$h$ to obtain the MAE and RMSE.
Furthermore, we evaluate 
\begin{align}
    \text{RatioOverprediction}^q_{h,g} &= \frac{1}{|\mathcal{G}|}\sum_{g \in \mathcal{G}}\mathbb{I}(\widehat{Q}^q_{h,g} - Q^q_{h,g} > 0)  &&&
    \text{Direction}^q_{h,g} &= \frac{1}{|\mathcal{G}|}\sum_{{g \in \mathcal{G}}}^T\operatorname{sign}(\widehat{Q}^q_{h,g} - \widehat{Q}^q_{h,g-1}) \\
    \text{Stability}^q_{h,g} &= |\widehat{Q}^q_{h,g} - \widehat{Q}^q_{h,g-1}|\label{eq:stability}
\end{align}
as more forecast-utility related metrics. Again, these metrics are aggregated by averaging over the indices~$g, q$ or~$h$. 

%%% Section 5.2: Statistical Results
\subsection{Statistical Results}\label{sec:forecasting-results-results}
\begin{figure}[t]
    \centering
    \includegraphics[width=0.33\linewidth]{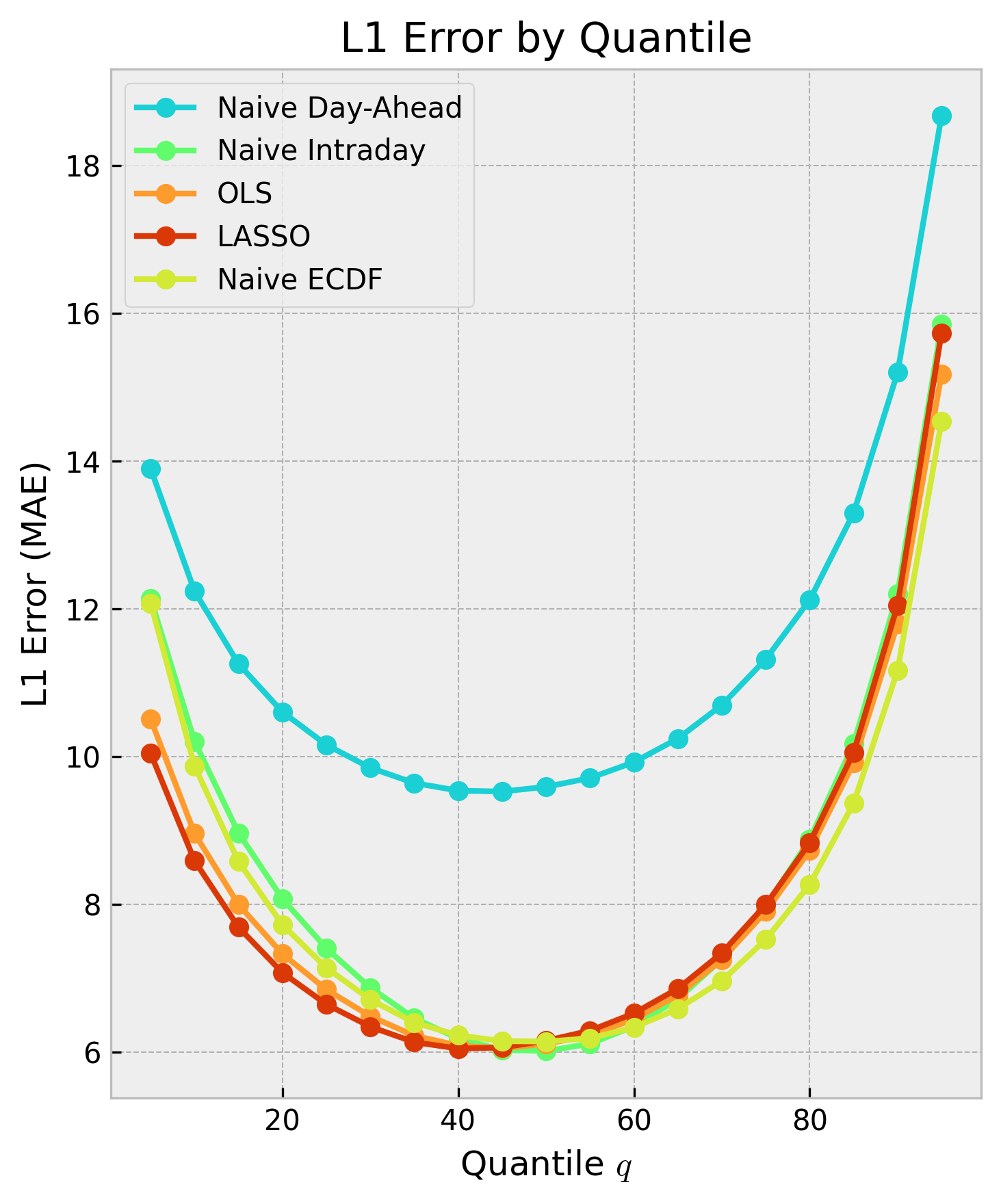}%
    \includegraphics[width=0.33\linewidth]{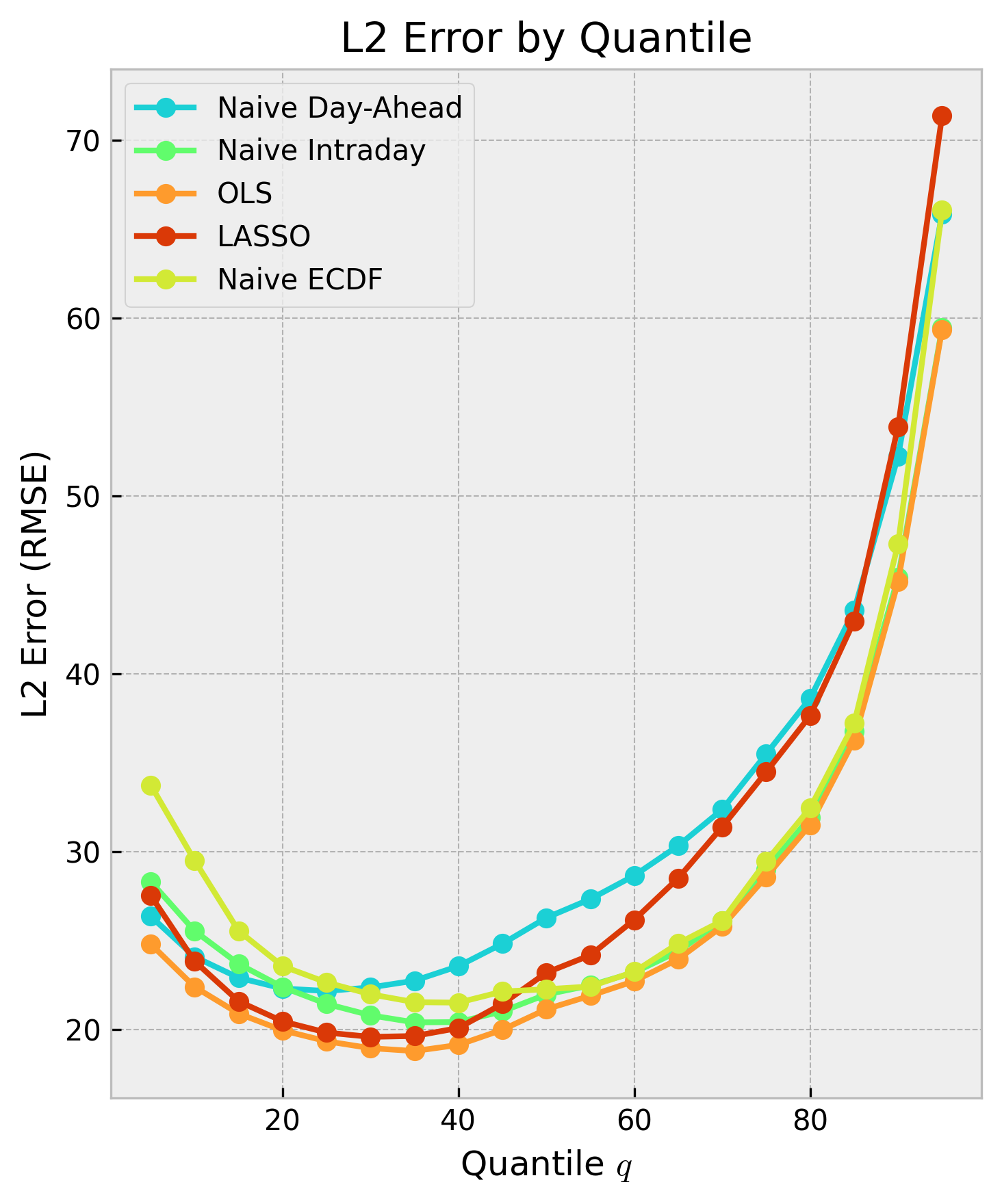}
    \caption{Error Metrics per Quantile Level $q$. $L_1$ and $L_2$ error metrics. }
    \label{fig:error_metrics_losses}
\end{figure}

Figure~\ref{fig:error_metrics_losses} presents the $L_1$ and $L_2$ error of the forecasts by quantile level. We observe that the error increases for lower and higher quantile levels. This is in line with expectations, as the tail quantiles are influenced by extreme values and are hence more difficult to predict. In contrast to \emph{probabilistic} forecasting, which focuses on estimating the uncertainty of a scalar variable, we focus on \emph{distributional} forecasting for the vector-valued empirical distribution. Importantly, this result does not contradict the existing literature on \emph{probabilistic} forecasting. While prior work primarily improves uncertainty quantification for extreme events \citep[see e.g.][]{hirsch2024simulation,narajewski2020ensemble,nickelsen2025bayesian}, our model specifically predicts the \emph{realized} values of those extreme tail quantiles.

\begin{figure}[htb]
    \centering
    \includegraphics[width=0.5\linewidth]{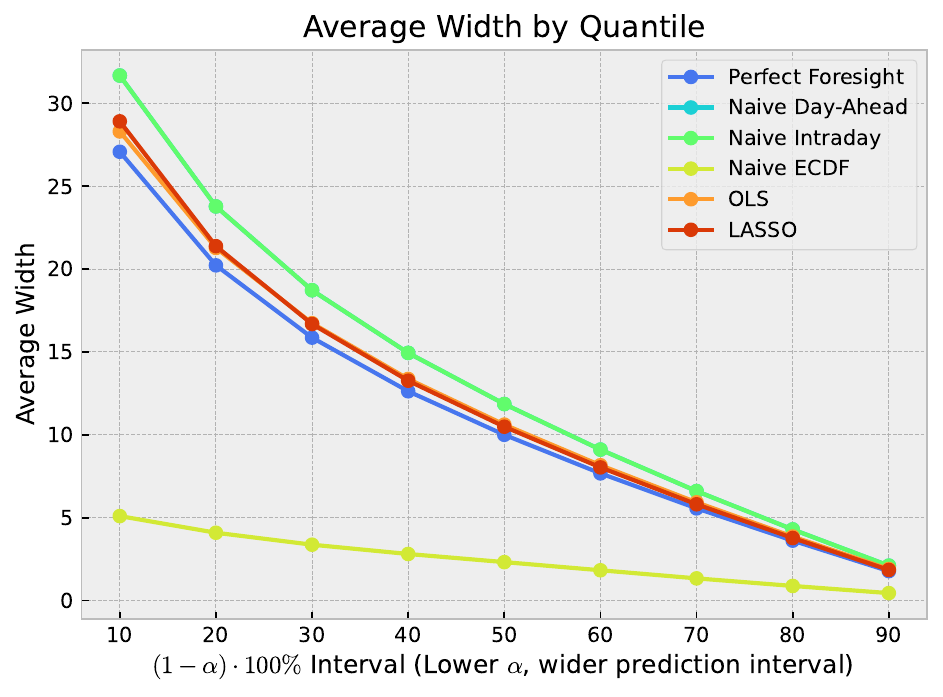}%
    \includegraphics[width=0.5\linewidth]{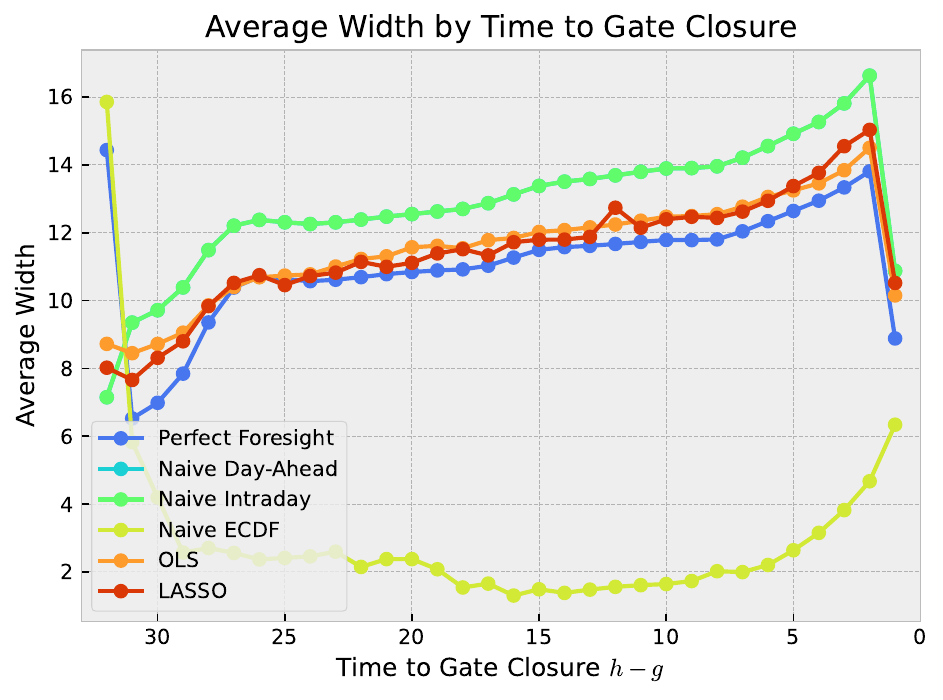}
    \caption{Average prediction interval width by interval width $\alpha$ and time to gate closure.}
    \label{fig:error_metrics_width}
\end{figure}

Figure~\ref{fig:error_metrics_width} gives the average width of the prediction intervals by quantile level and time to delivery. Generally, the perfect foresight, naive forecasts, and the \textsf{OLS} and \textsf{LASSO} models are close together, only the benchmark forecast based on the current trading session's ECDF provides significantly narrower prediction intervals. The width of the prediction intervals increases as we get closer to delivery, with a sharp drop in the last hour, which aligns with the intuition that the price distribution becomes more volatile as we approach delivery. It is well-known that the volatility of the intraday market is largest in the last 1-3 hours before the start of physical delivery \citep{narajewski2020ensemble,serafin2022trading, hirsch2024simulation,chen2025probabilistic}. Due to the construction of the target variable, which takes all mid prices in the remaining trading session into account, we expect that for early forecasts, the effect of volatility is somewhat dampened due to the longer remaining session. Once the most volatile period is over, the width of the prediction intervals drops significantly.

\begin{figure}[htb]
    \centering
    \includegraphics[width=0.33\linewidth]{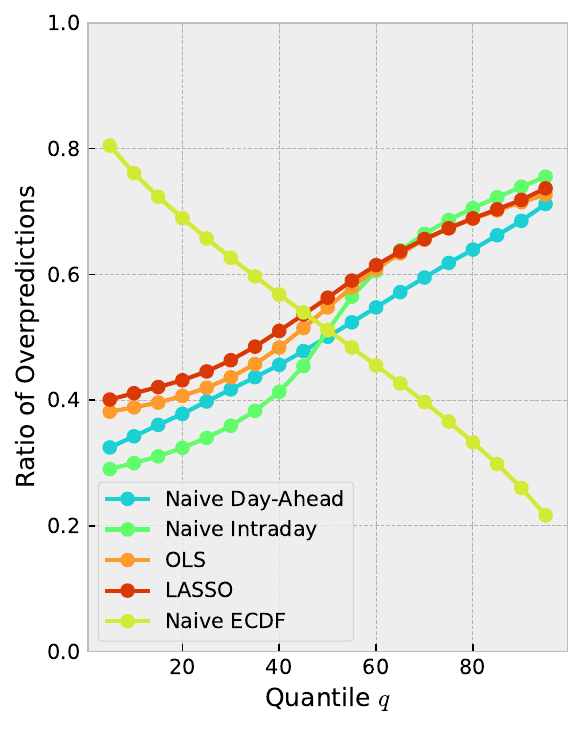}%
    \includegraphics[width=0.33\linewidth]{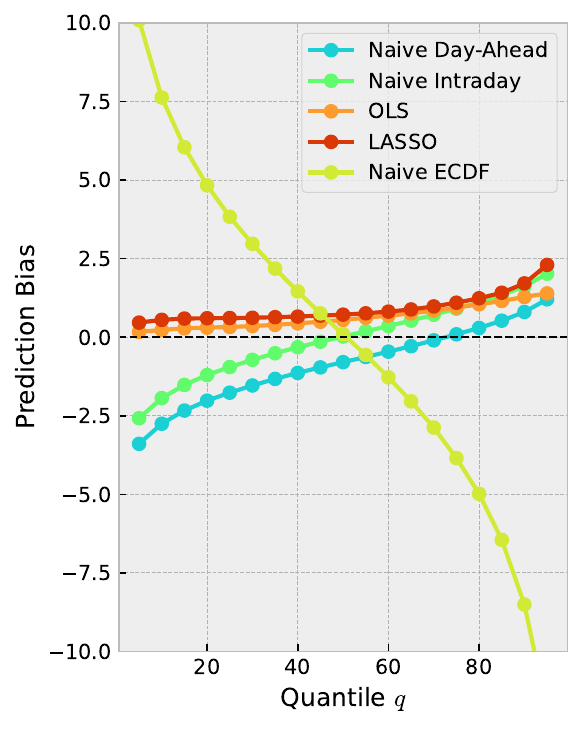}%
    \includegraphics[width=0.33\linewidth]{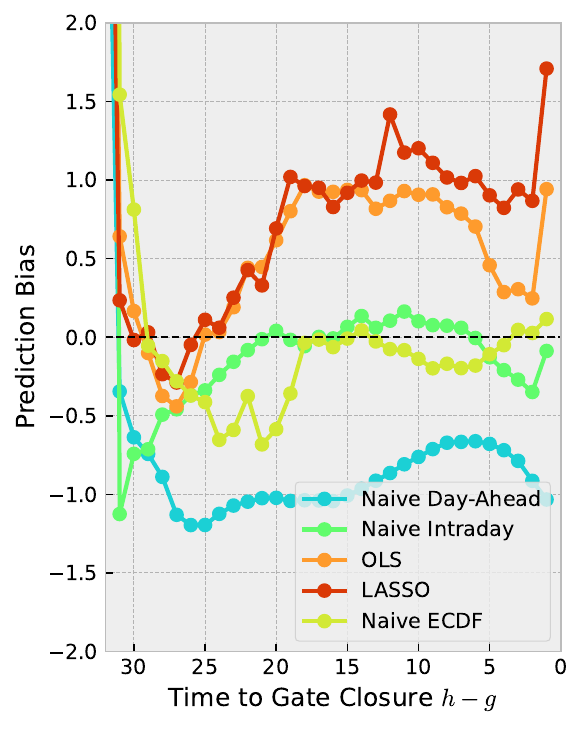}
    \caption{Bias by quantile level $q$ and time to gate closure.}
    \label{fig:error_metrics_bias}
\end{figure}

\begin{figure}[htb]
    \centering
    \includegraphics[width=0.5\linewidth]{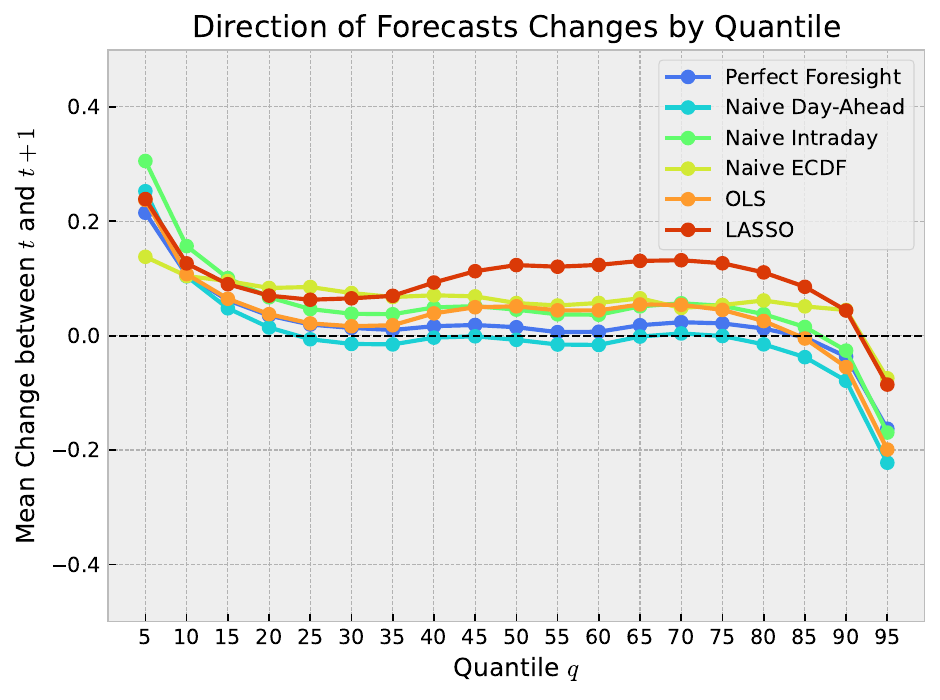}%
    \includegraphics[width=0.5\linewidth]{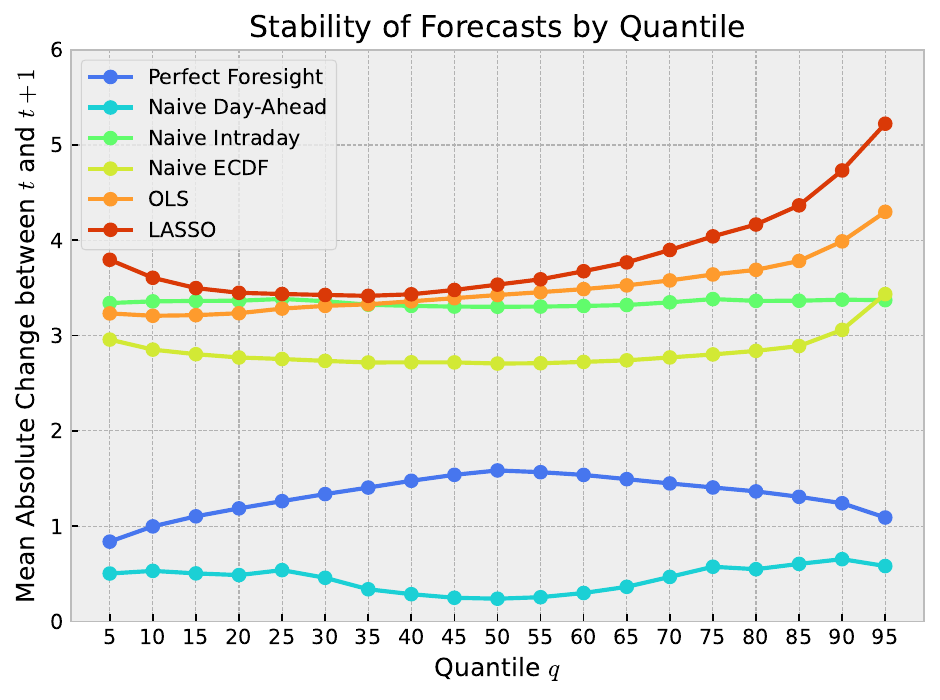}
    \caption{Directional changes and forecast stability by quantile. Forecasts with low quantiles increase, forecasts for high quantiles decrease. }
    \label{fig:error_metrics_direction_updates}
\end{figure}

Figure~\ref{fig:error_metrics_bias} shows the bias of the forecasts by quantile level and time to delivery. We generally observe that the low quantiles are biased downwards, while the high quantiles are biased upwards. This implies that there is less actual price dispersion than predicted by the models, and consequently, less liquidity outside of a potential $(1-\alpha)$ prediction interval. For the \textsf{NaiveECDF} benchmark, the effect is reversed. We generally see that the models have a small positive bias over the trading session, driven by the larger quantiles (middle and right panel of Figure~\ref{fig:error_metrics_bias}). Note, however, that the bias is small relative to the range of electricity prices.

Figure~\ref{fig:error_metrics_direction_updates} presents the directional changes and stability of the forecasts by quantile level. We see that for low quantiles, the forecasts tend to update upwards, while for high quantiles, the forecasts tend to update downwards. In contrast, Figure~\ref{fig:error_metrics_width} shows that the average width seems to increase throughout the trading session, with a sharp drop in the last hour. However, it is important to keep in mind that not all trading sessions span the full 32 hours to gate closure and thus the average across the different times to delivery bar increasing weight. The right panel of Figure~\ref{fig:error_metrics_direction_updates} indicates that the forecasts for the low and high quantiles are less stable than those for the middle quantiles, which aligns with the intuition that the tail quantiles are influenced by extreme values and are therefore more difficult to predict. We also observe that the \textsf{LASSO-based} model provides the lowest stability for the high quantiles.  Note that these plots display the quantile level $q$ instead of the interval parameter $\alpha$, as this distinction allows us to separate effects for the lower and upper tails of the distribution; however, for the width, we can only evaluate metrics based on the interval parameter $\alpha$.

All models show a similar stability level around the center, consistent with the assumption that the last traded price is a good predictor for the center of the price distribution, and hence little change between \textsf{NaiveECDF}, \textsf{OLS} and \textsf{LASSO} models is expected here. The \textsf{NaiveDayAhead} benchmark, by construction, has no updates in the level of the baseline price and only changes through the average ECDF of the rolling training session. Therefore, it generates highly stable forecasts. The \textsf{PerfectForesight}, on the other hand, has the largest updates and therefore the least stable predictions in the center of the distribution, even though the average absolute forecast update of the \textsf{PerfectForesight} is still below that of the \textsf{NaiveECDF} or the \textsf{NaiveIntraday}.

%%%%%%%%%%%%%%%%%%%%%%%%%%%%%%%%%%%%%%%%%%
%%% Section 6: Battery Trading Results %%%
%%%%%%%%%%%%%%%%%%%%%%%%%%%%%%%%%%%%%%%%%%
\section{Battery Trading Results}\label{sec:simulation-results}
In this section, we present the results of the economic evaluation of the forecasts for battery trading. The following section introduces the evaluation setting. Subsequently, we present results based on unconstrained trading experiments, where we use the forecasts to place limit orders at the predicted quantiles and evaluate the profits and trading behavior. We then present results based on an adaptive strategy, where we dynamically select the optimal parameters $\alpha$ and the optimal stop-loss parameter on a rolling calibration set (see Section \ref{sec:simulation-results-adaptive}) and evaluate the profits and trading behavior. We place special emphasis on the relationship between the different forecast metrics and economic performance. 

%%% Section 6.1: Setting
\subsection{Setting}\label{sec:simulation-results-setting}
Several optimization and simulation settings affect our trading behavior independently of the chosen forecasting method. For readability and structural clarity, we provide a complete overview of all quantitative simulation parameters in Appendix Table \ref{tab:sim_setting}. We keep all baseline technical and market parameters constant across our experiments to ensure fair comparability across models, varying only the battery energy storage capacity (MWh) and duration (hour = capacity/power) in our primary financial evaluations. A detailed overview of all technical battery parameters can be found in Appendix~\ref{sec:simulation-setting}.

Our empirical backtesting environment models the battery energy storage system across four distinct energy capacities of 10, 25, 50, and 100 MWh, paired with operational durations of one, two, and four hours. The simulations are based on continuous intraday LOB data from EPEX SPOT for the years 2023 and 2024.  We assume symmetric charging and discharging efficiencies of 95\%, accurately reflecting modern grid-scale lithium-ion battery installations. To capture market fees and prevent physical wear, we incorporate realistic EPEX exchange trading fees of 0.09 EUR/MWh alongside a linear degradation cost approximation of 4 EUR/MWh, following the baseline parameterization established by \citet{schaurecker2025maximizing}. All limit order volumes are multiples of 0.1 MWh as required by the exchange, and trading stops precisely 30 minutes before a product's physical delivery, when the market trading splits into the four German TSO zones.

We solve the forecast augmented intrinsic using the dynamic programming approximation introduced in \cite{schaurecker2025maximizing} in order to be able to match the millisecond speed of the market\footnote{See: https://github.com/dschaurecker/bitepy.}. For the optimization mechanics, our dynamic programming approximation discretizes storage-states into 11 discrete states, evaluating an action grid in increments of 10\% of the maximum storage to maintain computational tractability across two full historical simulation years. Each solve uses the latest available forecast to place virtual orders in the limit order book. Specifically, at solve time $t \in [g, g+1)$ for $g \in \mathcal{G}$ and a given parameter $\alpha$, we assign the forecast buy and sell prices using the tail quantiles $q_l = \alpha / 2$ and $q_u = 1 - \alpha / 2$, such that $P_{t,h}^{f,+} = \widehat{Q}^{q_l}_{h,g}$ and $P_{t,h}^{f,-} = \widehat{Q}^{q_u}_{h,g}$. Furthermore, to smooth out final scheduling decisions and avoid sudden, potentially costly position corrections shortly before delivery, we implement our fading mechanism \eqref{eq:fc_fading}, such that orders placed against the forecast start fading 120 minutes before delivery and completely phase out 40 minutes before delivery.

We benchmark our models both with and without an adaptive stop-loss. For the unconstrained models without the stop-loss, we run the battery optimization regardless of the drift between the forecast and the actually observed prices. For the simulations with stop-loss, we implement an adaptive strategy that performs a daily rolling hyperparameter search using the last 30-days as a calibration window $\mathcal{D}_\text{calib}$. For each upcoming trading day, this adaptive layer evaluates all combinations of prediction interval widths $\alpha$ and stop-loss risk aversion thresholds $\gamma \in \Gamma = \{0.5, 0.75, 0.9, 0.95, 0.99, 0.999\}$, dynamically selecting the optimal parameter pair $(\alpha, \gamma)$ based on cumulative profitability over $\mathcal{D}_\text{calib}$. Here, the threshold values $\gamma$ correspond to empirical quantiles of the weekly rolling intrinsic battery profits (without forecasts) across the preceding 364 days. 

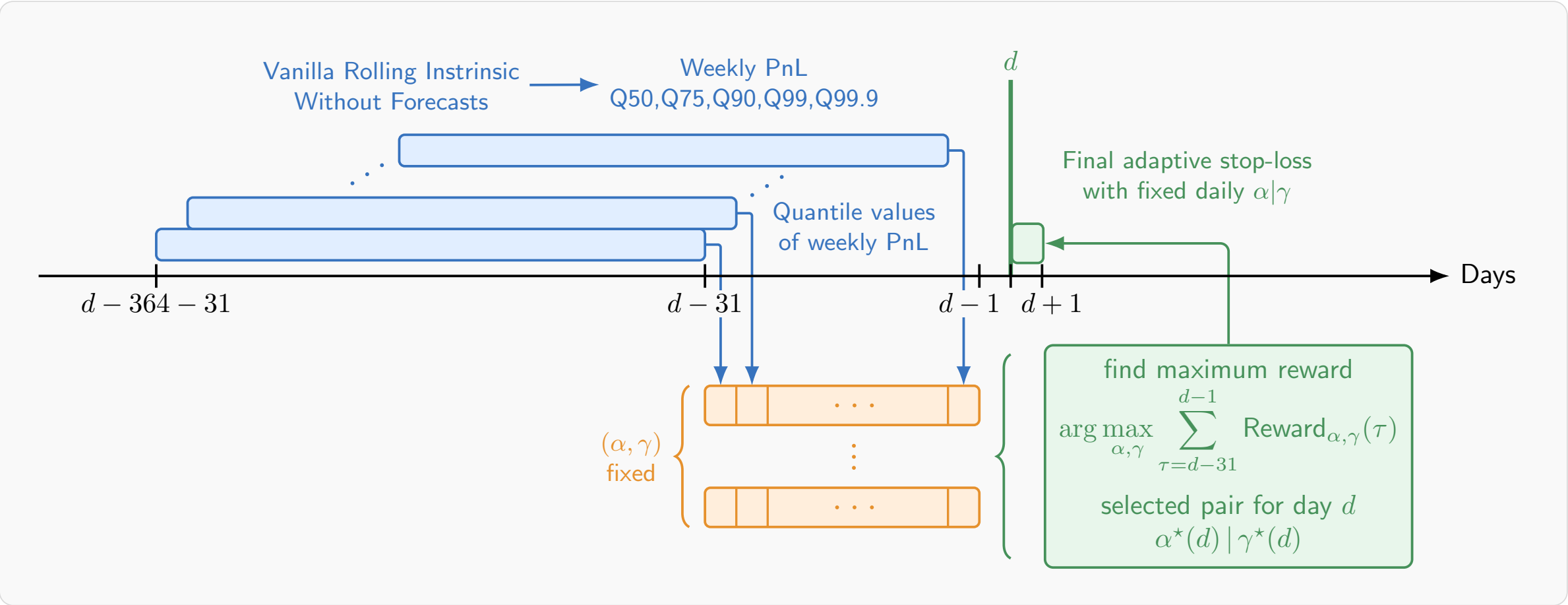
\begin{figure}[t]
    \centering
    \resizebox{0.9\textwidth}{!}{%
        \begin{tikzpicture}[
    font=\sffamily,
    >=Latex,
    box/.style={rounded corners=3pt, draw, fill=white, align=center, inner sep=5pt},
    bluebox/.style={box, draw=histblue, fill=softblue},
    orangebox/.style={box, draw=adaptorange, fill=softorange},
    greenbox/.style={box, draw=todaygreen, fill=softgreen},
    arrow/.style={->, line width=0.9pt},
    note/.style={font=\sffamily\small, align=center}
]

% 1. Background Box (Draw this first so it is behind everything)
\draw[draw=gray!30, fill=gray!5, rounded corners=5pt] (-2.0, -4.2) rectangle (18.0, 3.5);

% Time coordinates
\def\tA{0}      
\def\tB{7}      
\def\tC{10.5}   
\def\tD{10.9}
\def\tE{11.3} 
\def\tT{13.5}    

% Stacked Blue calibration windows
\node[note, text=histblue, anchor=south] (vanilla) at (3.0, 2.0) {Vanilla Rolling Instrinsic\\Without Forecasts};
\node[note, text=histblue, anchor=south] (newtext) at (7.5, 2.0) {Weekly PnL\\Q50,Q75,Q90,Q99,Q99.9};
\draw[arrow, histblue] (vanilla) -- (newtext);

% W1 
\draw[histblue, fill=softblue, rounded corners=2pt, line width=0.9pt] (0, 0.2) rectangle (7.0, 0.6);
\draw[arrow, histblue, rounded corners=1pt] (7.0, 0.4) -- (7.2, 0.4) -- (7.2, -1.4);

% W2 
\draw[histblue, fill=softblue, rounded corners=2pt, line width=0.9pt] (0.4, 0.6) rectangle (7.4, 1.0);
\draw[arrow, histblue, rounded corners=1pt] (7.4, 0.8) -- (7.6, 0.8) -- (7.6, -1.4);

% Continuation dots
\node[text=histblue, font=\Large, rotate=60] at (2.6, 1.35) {$\ddots$};
\node[text=histblue, font=\Large, rotate=60] at (7.7, 1.2) {$\ddots$};

% Wlast 
\draw[histblue, fill=softblue, rounded corners=2pt, line width=0.9pt] (3.1, 1.4) rectangle (10.1, 1.8);
\draw[arrow, histblue, rounded corners=1pt] (10.1, 1.6) -- (10.3, 1.6) -- (10.3, -1.4);

% Left brace for orange/yellow boxes
\draw[decorate, decoration={brace, mirror, amplitude=5pt}, adaptorange, line width=0.9pt]
    (6.8, -1.4) -- (6.8, -3.2)
    node[midway, left=6pt, note, text=adaptorange] {$(\alpha, \gamma)$\\fixed};

% Orange contiguous small boxes block with rounded outer edges (Top)
\draw[adaptorange, fill=softorange, rounded corners=2pt, line width=0.9pt] (7.0, -1.9) rectangle (10.5, -1.4);
\draw[adaptorange, line width=0.8pt] (7.4, -1.9) -- (7.4, -1.4);
\draw[adaptorange, line width=0.8pt] (7.8, -1.9) -- (7.8, -1.4);
\draw[adaptorange, line width=0.8pt] (10.1, -1.9) -- (10.1, -1.4);

% Orange horizontal dots inside the first small boxes block
\node[text=adaptorange, font=\Large] at (8.95, -1.65) {$\dots$};

% Vertical dots between the orange blocks
\node[text=adaptorange, font=\Large] at (8.9, -2.2) {$\vdots$};

% Orange contiguous small boxes block with rounded outer edges (Bottom)
\draw[adaptorange, fill=softorange, rounded corners=2pt, line width=0.9pt] (7.0, -3.2) rectangle (10.5, -2.7);
\draw[adaptorange, line width=0.8pt] (7.4, -3.2) -- (7.4, -2.7);
\draw[adaptorange, line width=0.8pt] (7.8, -3.2) -- (7.8, -2.7);
\draw[adaptorange, line width=0.8pt] (10.1, -3.2) -- (10.1, -2.7);

% Orange horizontal dots inside the second small boxes block
\node[text=adaptorange, font=\Large] at (8.95, -2.95) {$\dots$};

% Brace and explanation node (Now Green)
\draw[decorate, decoration={brace, amplitude=5pt}, todaygreen, line width=0.9pt]
    (10.9, -3.6) -- (10.9, -1.0)
    node[midway, right=12pt, greenbox, align=center] (biggreenbox) {
        find maximum reward\\[2pt]
        $\displaystyle \arg\max_{\alpha,\gamma}
        \sum_{\tau=d-31}^{d-1} \text{Reward}_{\alpha,\gamma}(\tau)$\\[6pt]
        selected pair for day $d$\\
        $\alpha^\star(d)\,|\,\gamma^\star(d)$
    };

% Small daily green box right next to t
\node[draw=todaygreen, fill=softgreen, rounded corners=2pt, line width=0.9pt, minimum width=0.4cm, minimum height=0.5cm, inner sep=0pt, anchor=south west] (smallgreen) at (\tD, 0.15) {};

% Green text right/top of the small green box
\node[note, text=todaygreen, anchor=south west] at ($(smallgreen.north east) + (0.1, 0.1)$) {Final adaptive stop-loss\\with fixed daily $\alpha \vert \gamma$};

% Green arrow from big green box to small green box
\draw[arrow, todaygreen, rounded corners=3pt] (biggreenbox.north) |- (smallgreen.east);
\node[note, text=histblue] at (8.9, 0.6) {Quantile values\\of weekly PnL};
% \node[note, text=adaptorange] at (8.75, -3.8) {Every ($\alpha \vert Q$) pair\\fixed for 30 days};

% 2. Foreground Elements (Drawn last to guarantee top z-order alignment)
% Vertical green line for day t
\draw[todaygreen, line width=1.6pt] (\tD, 2.5) -- (\tD, 0.0);
\node[above, text=todaygreen, font=\bfseries] at (\tD, 2.5) {$d$};

% Timeline and Axis Labels
\draw[->, line width=0.9pt] (-1.5, 0) -- (16.5, 0) node[right] {Days};

\draw[line width=0.9pt] (\tA, 0.15) -- (\tA, -0.15);
\node[below=5pt, fill=gray!5, inner sep=1pt] at (\tA, 0) {$d-364-31$};

\draw[line width=0.9pt] (\tB, 0.15) -- (\tB, -0.15);
\node[below=5pt, fill=gray!5, inner sep=1pt] at (\tB, 0) {$d-31$};

\draw[line width=0.9pt] (\tC, 0.15) -- (\tC, -0.15);
\node[below=5pt, fill=gray!5, inner sep=1pt] at (\tC-0.125, 0) {$d-1$};

\draw[line width=0.9pt] (\tD, 0.15) -- (\tD, -0.15);

\draw[line width=0.9pt] (\tE, 0.15) -- (\tE, -0.15);
\node[below=5pt, fill=gray!5, inner sep=1pt] at (\tE+0.125, 0) {$d+1$};

\end{tikzpicture}%
    }
    \caption{Daily optimization of the stop-loss parameter.}
    \label{fig:optimization-stop-loss}
\end{figure}

Crucially, we enforce a strict one-day gap between the end of the rolling calibration window and the target delivery day $d$. Because continuous intraday trading for delivery on day $d+1$ officially opens at 15:00 on day $d$, incorporating afternoon price dynamics from day $d$ into the calibration set would introduce look-ahead bias and severe data leakage \citep{bailey2015statistical}. Figure~\ref{fig:optimization-stop-loss} visualizes the execution sequence of this adaptive simulation procedure.

Our adaptive study evaluates an out-of-sample test period from February 2023 through December 2024. For each rolling test day $d$, we first calculate vanilla (no-forecast) PnLs over the preceding 364-day window $[d-364-31, d-31)$ to establish candidate stop-loss thresholds $\gamma$. We then select the optimal parameter pair $(\alpha, \gamma)$ by evaluating cumulative profitability over the subsequent 30-day calibration window $\mathcal{D}_\text{calib} = [d-31, d-1)$.

%%% Section 6.2: Economic Evaluation based on Adaptive Strategy
\subsection{Economic Evaluation based on Adaptive Strategy}\label{sec:simulation-results-adaptive}
\begin{figure}[t]
    \centering
    \includegraphics[width=0.8\textwidth]{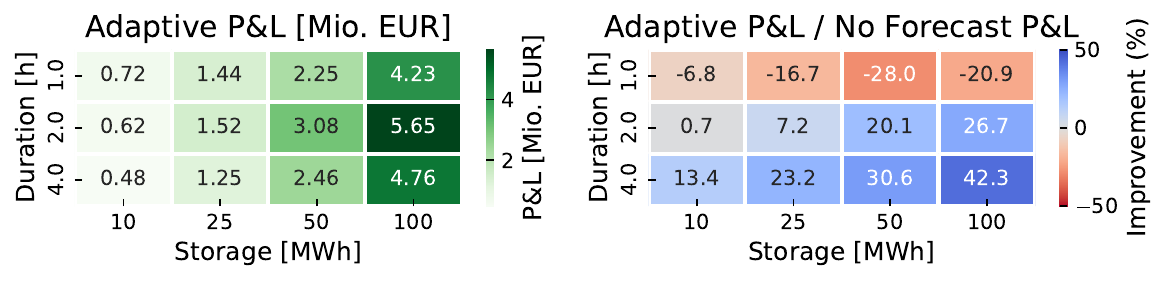}%
    \caption{Profits of the adaptive strategy relative to the benchmark strategy without forecasts. Percentage improvements are calculated as $100 \cdot (\frac{\text{Adaptive P\&{}L}}{\text{Benchmark P\&{}L}} - 1)$.}
    \label{fig:simulation_results_profits}
\end{figure}

In this section, we present the financial results of the adaptive trading strategy using the \textsf{LASSO} model. Our results show that the benefit of including the forecasts increases with battery size and duration. Figure~\ref{fig:simulation_results_profits} shows profits for the different durations and storage sizes. We present the total profits in mio.~EUR and the outperformance of the adaptive strategy over a benchmark strategy without forecasts. The adaptive strategy outperforms the benchmark strategy for the 2 and 4 hour duration batteries by up to 1\% to 27\% and 13\% to 42\% respectively. For the 1 hour battery, we see that the adaptive strategy underperforms the benchmark.

The increase in profits with larger battery capacity is likely due to the fact that forecast-driven trading strategies shift trading volumes to later trading periods, which avoids trading during low liquidity periods and disproportionately benefits larger storages, as liquidity costs increase superlinearly due to our LOB-based trading. The increase in profits with increasing duration is likely due to the fact that longer durations are "hedged" against single bad forecast decisions.

\begin{figure}[t]
    \centering
    \includegraphics[width=0.33\textwidth]{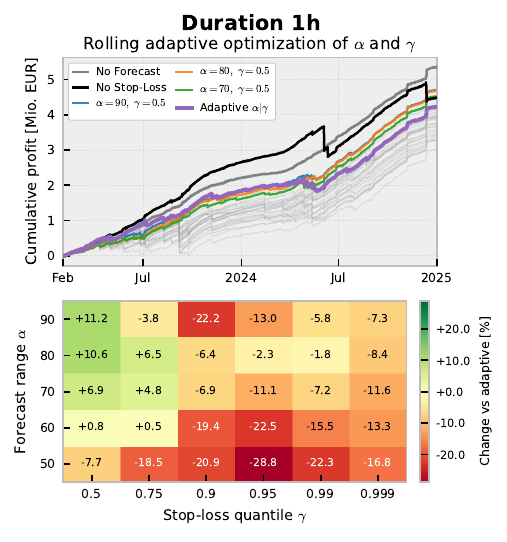}%
    \includegraphics[width=0.33\textwidth]{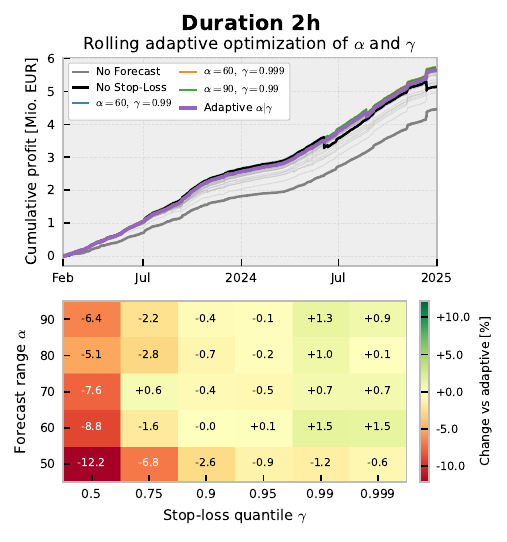}%
    \includegraphics[width=0.33\textwidth]{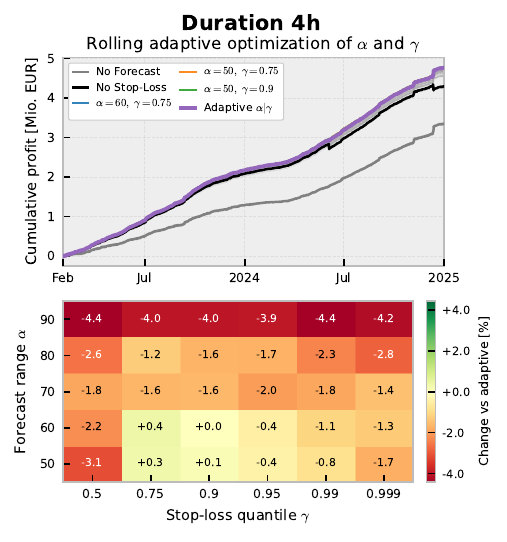}%
    \caption{Strategy runs underlying the adaptive strategy in for three configurations of the 100 MWh battery with duration 1, 2 and 4 hours. For the 1-hour duration battery, all $(\alpha, \gamma)$ parameter pairs yield lower profits than the no-forecast baseline, while omitting a stop-loss safeguard results in excessively costly schedule corrections. The three highlighted settings are the best-performing fixed settings of each duration. The lower heatmap presents each fixed-setting's performance against the adaptive strategy.}
    \label{fig:simulation_results_profits_100mwh_heatmap}
\end{figure}

We explore the differences in the increase concerning the duration in Figure~\ref{fig:simulation_results_profits_100mwh_heatmap}. We plot the underlying unconstrained experiments for the adaptive strategy for a 100 MWh battery across the 1, 2, and 4 hour durations. We see that for the 1-hour battery, the individual experiments have high variance and underperform the benchmark strategy and, thus, the joint adaptive selection is unable to find a combination of forecast and stop-loss parameters that outperforms the benchmark. It also becomes apparent that the 1-hour battery performs worse for riskier forecasts, i.e., smaller values of $\alpha$. For the 2- and 4 hour batteries, we see that the individual experiments have lower variance and generally outperform the benchmark strategy, enabling the adaptive strategy to find a good combination of forecast and stop-loss parameters. Although the adaptive strategy yields slightly lower revenue than the optimal fixed-parameter setting for each duration, largely due to premature stop-loss triggers from false alarms, it successfully eliminates the severe PnL drawdowns. Moreover, comparing it to the best fixed-parameter benchmark is unrealistic, as identifying the perfect configuration in advance is not possible.

%%% Section 6.3: Economic Evaluation based on Unconstrained Trading Experiments 
\subsection{Economic Evaluation based on Unconstrained Trading Experiments} \label{sec:simulation-results-naive}
\begin{figure}[htb]
    \centering
    \includegraphics[width=\textwidth]{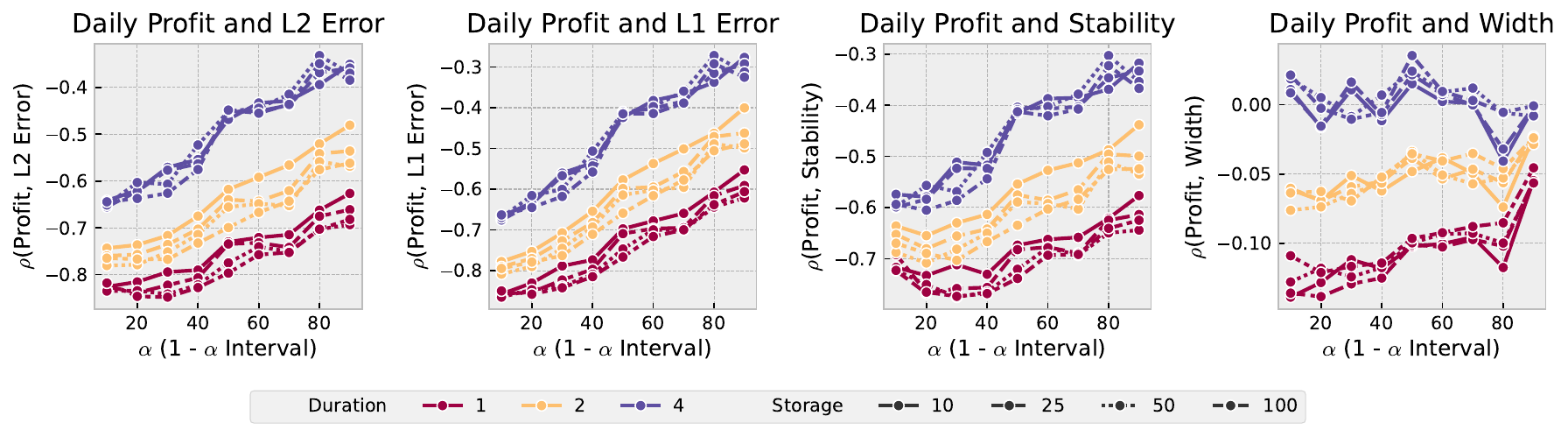}
    \caption{Correlation of daily profits and the daily ratio of forecast/market cashflow with forecast errors metrics. We use $1 - \alpha$ prediction intervals, hence lower $\alpha$ levels correspond to wider prediction intervals.}
    \label{fig:simulation_results_correlation}
\end{figure}

\begin{figure}[t]
    \centering
    \includegraphics[width=\textwidth]{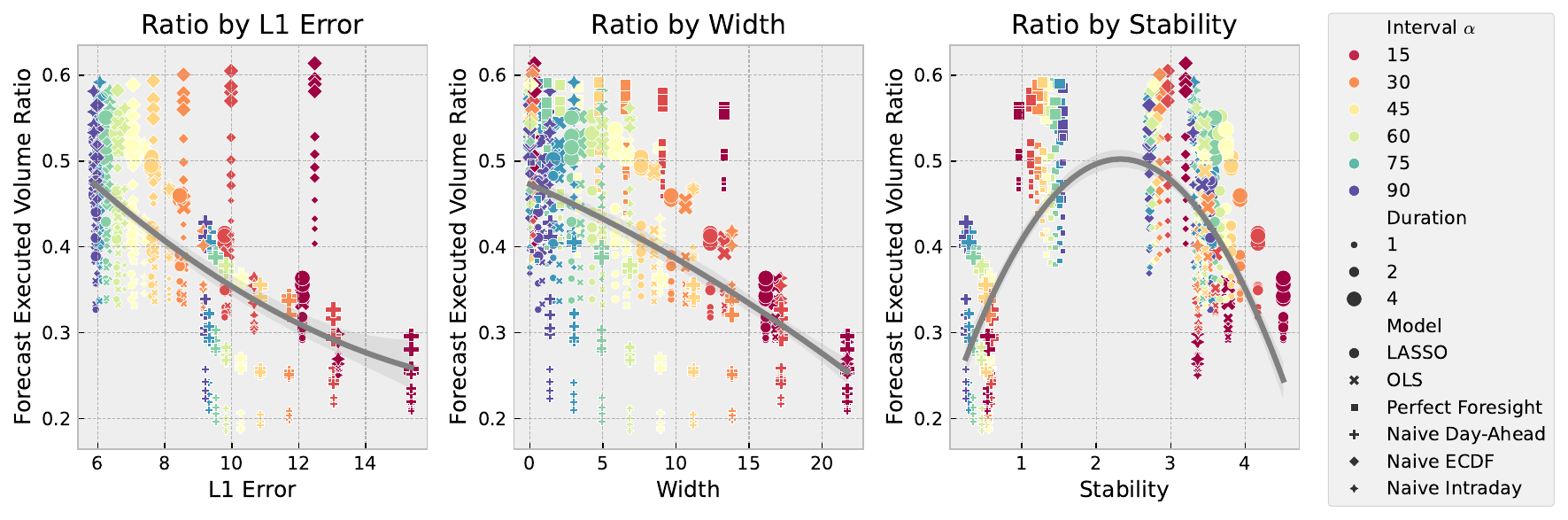}
    \caption{Forecast-based order share by forecast metrics, where the y-axis is the fraction of total gross executed volume traded through forecast-based orders over the total executed volume based on all submitted orders. The plot does not differentiate between the different battery sizes used in our case studies, we only differentiate between $\alpha$, duration and model, by color, marker size, and symbol.}
    \label{fig:simulation_results_forecast_ratios}
\end{figure}

\begin{figure}[t]
    \centering
    \includegraphics[width=\textwidth]{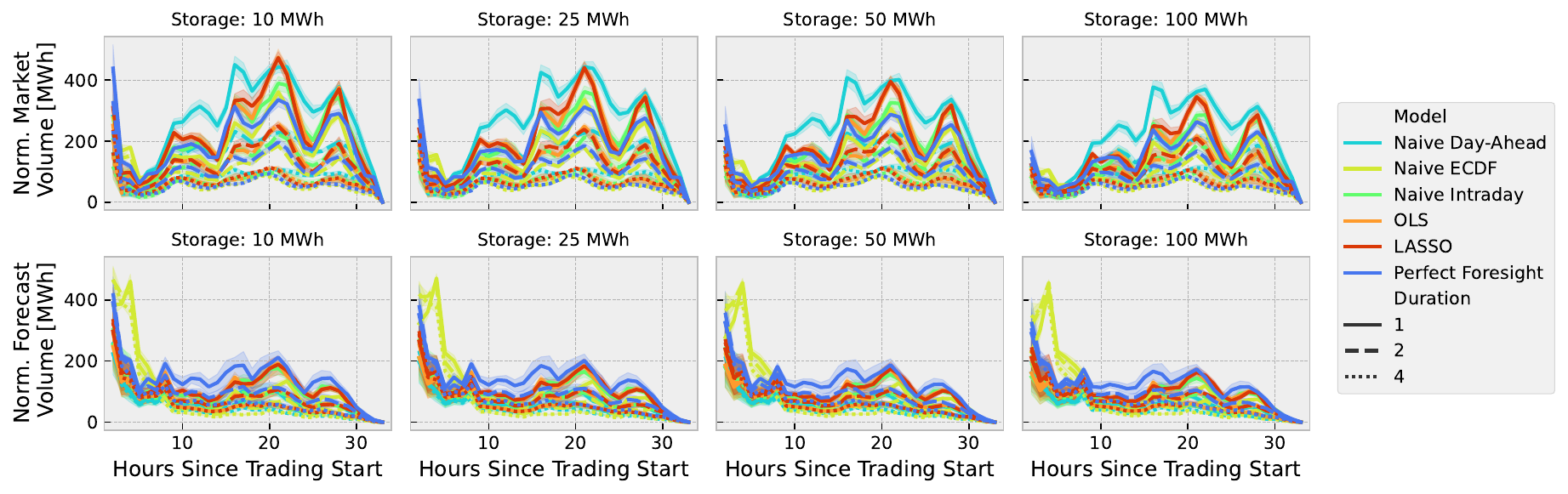}%
    \caption{Market-order traded volumes and forecast-order traded volumes over the daily trading session. Volumes are normalized by the installed storage capacity. We take the sum of all executed trades, grouped by an hourly grid $g$ along the trading time and normalize the volumes by the installed storage capacity for better comparison across battery configurations.}
    \label{fig:market-forecast-traded-volumes-session}
\end{figure}

\begin{figure}[t]
    \centering
    \includegraphics[width=\textwidth]{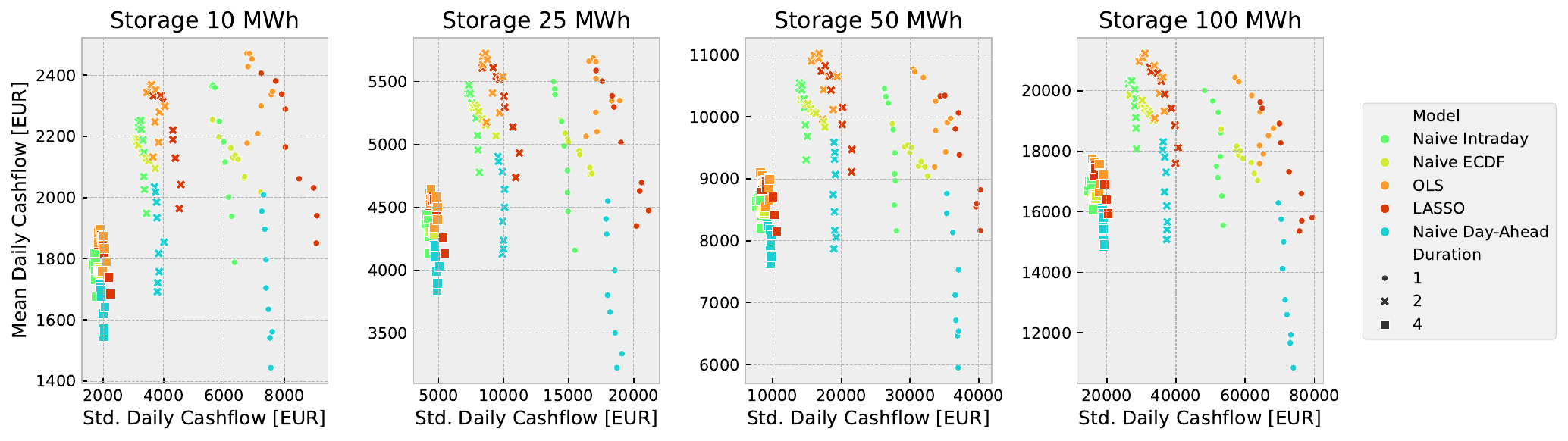}
    \caption{Risk-return profiles of the different forecasting models.}
    \label{fig:risk-return-profiles}
\end{figure}

In this section, we present the financial results of the different combinations of forecast models, prediction interval widths $\alpha$, and battery sizes and durations, omitting any stop-loss to focus on the underlying dynamics. This collection provides us with a cross-section of different forecast qualities and the associated trading behaviors. We focus on the relationship between forecast quality and economic performance. Our main insights can be summarized as follows: \begin{itemize}
    \item Forecast errors are negatively correlated with profits, that is, higher errors lead to lower profits. This is in line with the intuition that better forecasts lead to better trading decisions and, hence, higher profits. 
    
    \item Instable forecasts are often linked to rebalancing costs \citep{genov2025balancing,godahewa2025forecast} and our results further confirm this link. Again, we find that for for smaller $\alpha$, the correlation is stronger. 
    
    \item The share of executed forecast-based limit orders increases with forecast quality and decreases with forecast interval width. Wider prediction intervals lead to more passive placements of limit orders deeper in the order book; hence, forecast orders are less likely to be executed.
\end{itemize}
We discuss the details in the following paragraphs and Figures \ref{fig:simulation_results_correlation} to \ref{fig:market-forecast-cashflow}.

Figure~\ref{fig:simulation_results_correlation} shows the correlations between daily profits and forecast error metrics. $L_1$ and $L_2$ forecast errors are negatively correlated with profits, that is higher errors lead to lower profits. This aligns with the intuition that better forecasts lead to better trading decisions and, hence, higher profits. The correlation is stronger for smaller $\alpha$, which supports the notion that smaller $\alpha$ leads to more aggressive trading strategies that are more sensitive to forecast errors. 

This result is in contrast to \cite{maciejowska2026statistical, serafin2025data}, who conduct a correlation analysis among many day-ahead forecasting models and their associated profits on a sample level and find only modest levels of correlation between forecast errors and profits. They conclude that traditional $L_1$ and $L_2$ metrics have no strong association with profits. Our results show that on a daily and thus operational horizon, the relationship between $L_1$ and $L_2$ error metrics is clearly visible, indicating that our forecasts are useful for the augmented RI.

A similar pattern holds for the stability of forecasts, where we note that stability is negatively defined and therefore higher values indicate less stable forecasts. Instable forecasts are often linked to switching costs if decisions can only be reversed at high costs \citep{genov2025balancing,godahewa2025forecast} and our results further confirm this link. Again, we find that for smaller $\alpha$, the correlation is stronger, which is plausible as reversing decisions based on the tails of the price distribution is more costly than decisions based in the center of the distribution. Lastly, we see no strong patterns for the impact of the width of the prediction intervals on profits.

Figure \ref{fig:simulation_results_forecast_ratios} shows the ratio between the volume traded through direct-orders and the volume traded through forecast-based limit orders, as well as the volume traded directly on the market. We see that the share of volume traded via forecast-based limit orders increases with the quality of the forecast and with decreasing $\alpha$. Wider prediction intervals lead to more passive placement of limit orders deeper in the order book. Hence, forecast orders are less likely to be executed—if they are, it is at a higher profit. In the right-most panel, we see that the link between stability and the volume ratio has an inverse \emph{U-shape}.
% TODO: More drivers?

We recollect the correlations presented in Figure~\ref{fig:simulation_results_correlation} and see that larger intervals are linked to higher forecast errors and lower profits, but also to a smaller share of forecast-based limit orders, and similar for the width of the forecasts. Comparing the first and third panels, we see that the $\textsf{NaiveDayAhead}$ (denoted with the \textbf{+} symbol) brings high stability at the cost of high $L_1$ forecast errors, emphasizing that such metrics should not be evaluated in isolation. 

Figure \ref{fig:market-forecast-traded-volumes-session} shows the traded volumes through direct-orders and forecast-based limit orders during the trading session. We see that forecast orders are more likely to be traded in the first hours of the trading session, while direct-orders are more likely to be traded in the remaining hours. Note that due to the different gate-closure times of the various delivery periods, the shortest trading session lasts only 8 hours, and the longest trading session lasts 31 hours. Hence, we expect an increase in direct-orders to close positions around 7 hours after the start of trading, in line with the top row of Figure \ref{fig:market-forecast-traded-volumes-session}. We also observe distinct peaks corresponding to the gate-closure times of the morning and afternoon delivery periods. As expected, the \textsf{PerfectForesight} experiment has a higher trading volume for forecast-based orders also in the later trading sessions due to its ability to perfectly capture profitable trading opportunities. The \textsf{NaiveDayAhead} benchmark shows a higher direct-order volume in the initial hours of the trading session. This might be based on the execution of initial trades as the intraday market might start at a different price level than the day-ahead price and hence, the centering around the day-ahead prices is suboptimal. 

We show the risk-return profiles of the different forecasting models in Figure~\ref{fig:risk-return-profiles}. Generally, we see that risk-return patterns are largely driven by the forecast model and are broadly consistent across storage capacities. Risk-return curves are steeper for longer durations; that is, we achieve higher returns for the same level of risk. Because longer-duration batteries must distribute their charging and discharging cycles over multiple hours, they inherently diversify their market exposure, flattening the risk-return curve by hedging against isolated single-hour forecast errors. Generally, the benchmark models achieve lower returns at lower risk levels, while the more advanced models achieve higher returns at higher risk levels. We see that risk-return profiles are generally slightly increasing in the interval size (not shown in Figure~\ref{fig:risk-return-profiles}), that is, wider intervals yield better risk-adjusted returns.

%%%%%%%%%%%%%%%%%%%%%%%%%%%%%
%%% Section 7: Conclusion %%%
%%%%%%%%%%%%%%%%%%%%%%%%%%%%%
\section{Conclusion}\label{sec:conclusion}
This paper introduces a novel approach for the integration of distributional forecasts in the optimization of battery trading in the intraday electricity market. Based on the structure of the intraday market and its unique challenges, such as the high-frequency nature of trading, uneven lengths of trading sessions, and order book dynamics, we develop a flexible, intuitive framework for the interface between optimization and forecasting. We apply this framework to a case study of battery trading in the German intraday market, using a range of forecasting models and optimization settings. Our results show that the inclusion of distributional forecasts can lead to significant improvements in trading performance, especially for larger battery sizes and longer durations. We also find that forecast errors and stability are important factors influencing the economic performance of the trading strategy and that the share of forecast-based limit orders increases with forecast quality and decreasing prediction interval width.

While our results provide valuable insights into the integration of distributional forecasts in battery trading optimization, there are several limitations to our study that should be acknowledged. These limitations also point towards interesting avenues for future research. First, our analysis is limited to hourly products, and an extension to include quarter-hourly products would be interesting, although it would introduce additional challenges, such as the curse of dimensionality for longer durations. Second, we focus on relatively simple, albeit established, forecasting models, such as expert-type OLS and LASSO models. Exploring more advanced machine learning models could potentially lead to further improvements in forecast quality and trading performance. Third, our backtesting approach does not account for market reactions to our trades, which could affect the profitability of the strategy in a real-world setting. Finally, we use the myopic rolling intrinsic strategy. Moving towards a stochastic optimization framework that directly incorporates probabilistic forecasts could be a promising direction for future research.

\section*{Acknowledgments}
Simon Hirsch is employed as an industrial PhD student by Statkraft Trading GmbH and gratefully acknowledges the support and funding received. 
The authors are grateful to Daniel Gruhlke and Florian Ziel for their helpful comments and suggestions on the paper.
The authors are grateful to Statkraft and an anonymous energy meteorology company for providing the data used in this paper.
This work contains the author's opinions and does not necessarily reflect Statkraft's position. The authors declare no conflicts of interest.

\section*{Generative AI Statement}

During the preparation of this work the authors used Github Copilot (multiple LLM models) in order to improve the quality of the code, figures, tables and text. After using this tool/service, the authors reviewed and edited the
content as needed and take full responsibility for the content of the published article.

\bibliography{refs.bib}

\appendix

\section{Simulation Settings}\label{sec:simulation-setting}

Table \ref{tab:sim_setting} specifies all simulation, optimization, and market settings used for our backtesting study referenced in Section \ref{sec:simulation-results-setting}.

\begin{table}[htb]
    \centering
    \adjustbox{width=0.8\textwidth}{
    \begin{tabular}{llccc}
        \toprule
        {Category} & {Parameter} & {Symbol} & {Value(s)} & {Unit} \\
        \midrule
        Battery Storage Specs & Energy Capacity & $\bar{s}$ & $ \{ 10,25,50,100 \} $ & MWh \\
                              & Battery Duration & -- & $\{ 1,2,4 \}$ & h \\
                              & Efficiency (Charge/Discharge) & $\eta^+, \eta^-$ & 0.95 & -- \\
        \midrule
        Costs \& CID Market   & Trading Fee & $\xi_{\text{trade}}$ & 0.09 & EUR/MWh \\
                              & Degradation Cost (Linear) & $\xi_{\text{deg}}$ & 4.00 & EUR/MWh \\
                              & Min. Tradable Unit & $u$ & 0.1 & MWh \\
                              & Gate Closure & -- & 30 & min \\
                              & Simulated Years & -- & 2023,2024 & -- \\
        \midrule
        Optimization Settings & Action Discretization & $\kappa$ & $\lfloor \bar{s} / 10 \rfloor$ & -- \\
                              & State Grid Size & $|G|$ & 11 & -- \\
        \midrule
        Forecasting Settings  & Forecast Fade Start & $T_{\text{fade}}^{\text{start}}$ & 120 & min \\
                              & Forecast Fade End & $T_{\text{fade}}^{\text{end}}$ & 40 & min \\
        \bottomrule
    \end{tabular}
    }%
    \caption{Baseline simulation parameters for the BESS, the intraday market environment, and the DP.}
    \label{tab:sim_setting}
\end{table}

\section{Benchmark Models}\label{app:benchmarks}

This section briefly describes the \textsf{NaiveDayAhead}, \textsf{NaiveIntraday} and the \textsf{NaiveECDF} benchmark models.For the \textsf{NaiveECDF} benchmark, we use the empirical cumulative distribution function of the trading session up to $g$ as forecasts, formally:
\begin{equation}
    \hat{Q}^{q,\textsf{NaiveECDF}}_{h,g} = \underbrace{
        F^{-1}_q\left(\left\{\dfrac{P^{[1,\text{bid}]}_{h,t} + P^{[1,\text{ask}]}_{h,t}}{2}\right\} \forall t \leq g \right)
    }_\text{Time-weighted ECDF of the trading session up to $g$.} 
\end{equation}
where $F^{-1}_q$ denotes the $q$-quantile of the time-weighted ECDF of the trading session up to $g$. The time-weighted ECDF is constructed as in Section \ref{sec:limit-order-book-features}. The \textsf{NaiveDayAhead} and \textsf{NaiveIntraday} benchmarks the target variable normalized by the day-ahead respectively intraday price as forecasts. Formally, we have:
\begin{equation}
    \hat{Q}^{q,\textsf{NaiveDayAhead}}_{h,g} = P_h^\text{DA} 
        + \frac{1}{363} 
        \sum_{d=2}^{364} \left(\hat{Q}^{q,}_{h(-d), g} - P_{h(-d)}^\text{DA}\right)
\end{equation}
and 
\begin{equation}
    \hat{Q}^{q,\textsf{NaiveIntraday}}_{h,g} = P_h^{[\text{mid},1h]} 
        + \frac{1}{363} 
        \sum_{d=2}^{364} \left(\hat{Q}^{q}_{h(-d), g} - P_{h(-d),g}^{[\text{mid},1h]}\right)
\end{equation}
and the notation $h(-d)$ denotes the same delivery period $h$ on the day $d$ days before the current trading day. We take two days of lag to account for the fact that the parallel trading sessions otherwise would introduce lookahead bias in the forecasts.
\section{Additional Figures and Tables}

\begin{figure}[htb]
    \centering
    \includegraphics[width=0.6\textwidth]{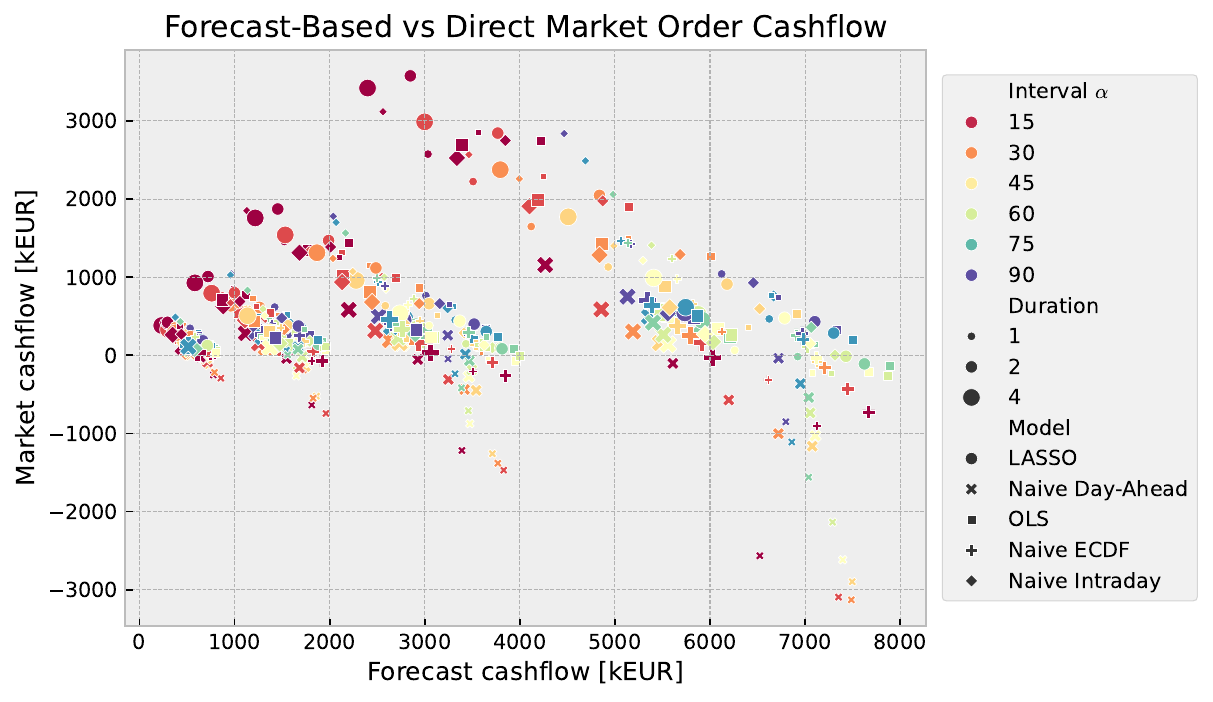}
    \caption{Direct-order cashflow and forecast-order cashflow by forecast model, duration and prediction interval width.}
    \label{fig:market-forecast-cashflow}
\end{figure}

We show the total cashflow of direct-orders and forecast-based limit orders by duration, model and prediction interval width in Figure~\ref{fig:market-forecast-cashflow}. We see that there is a large variation in the cashflow of forecast orders, with forecast-based orders generally inducing positive total cashflows, while direct-order cashflow are positive and negative across the configurations. Larger durations and wider prediction intervals are linked positive forecast-based total cashflows, but overall the variation is large and patterns not clear.

\begin{figure}[htb]
    \centering
    \includegraphics[width=\textwidth]{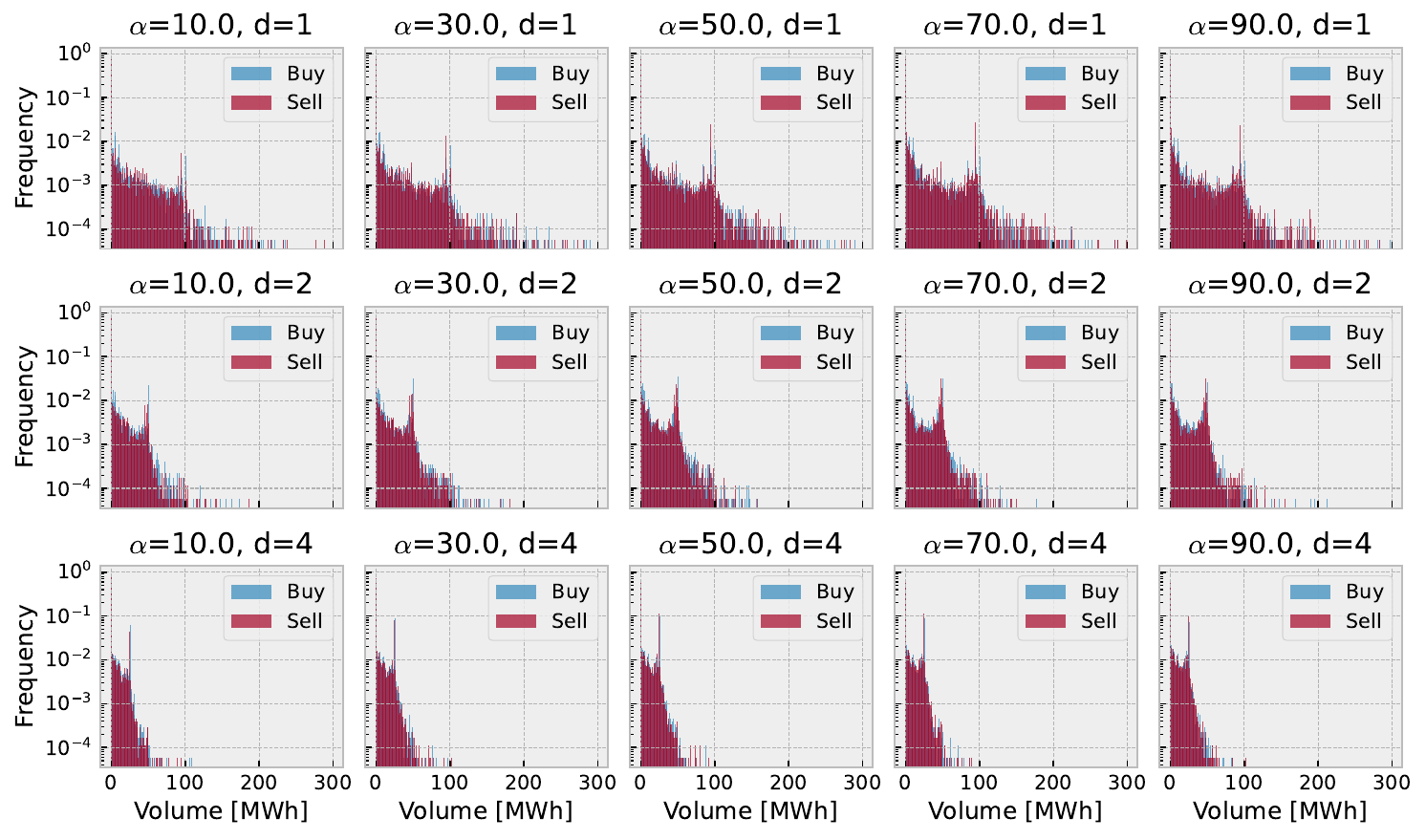}
    \caption{Histograms of traded volumes for forecast-based limit orders per trading session for the LASSO model, storage capacity of 100 MWh.}
    \label{fig:volume-histograms}
\end{figure}

We show histograms of the traded volumes for forecast-based limit orders per trading session for the LASSO model, storage capacity of 100 MWh in Figure~\ref{fig:volume-histograms}. We see that the distribution of traded volumes is right-skewed, with a large share of small volumes and a small share of large volumes. The distribution of traded volumes is more right-skewed for smaller prediction interval widths, which is in line with the intuition that smaller prediction intervals lead to more aggressive trading strategies, which are more likely to lead to larger profits but also larger losses. The distribution of buy and sell volumes volumes the forecast orders is similar, with a slightly higher share of larger volumes for sell orders.

\begin{table}[htb]  
    \centering
    \adjustbox{width=\textwidth}{%
\begin{tabular}{lrrrrrrrrrrrr}
\toprule
Battery Size (MWh) & \multicolumn{3}{c}{10} & \multicolumn{3}{c}{25} & \multicolumn{3}{c}{50} & \multicolumn{3}{c}{100} \\
Duration (h) & 1 & 2 & 4 & 1 & 2 & 4 & 1 & 2 & 4 & 1 & 2 & 4 \\
\midrule
No Forecast & {\cellcolor[HTML]{F7FBFF} \color[HTML]{000000} 577.54} & {\cellcolor[HTML]{F7FBFF} \color[HTML]{000000} 432.88} & {\cellcolor[HTML]{F7FBFF} \color[HTML]{000000} 282.13} & {\cellcolor[HTML]{F7FBFF} \color[HTML]{000000} 1,249.94} & {\cellcolor[HTML]{F7FBFF} \color[HTML]{000000} 946.30} & {\cellcolor[HTML]{F7FBFF} \color[HTML]{000000} 630.67} & {\cellcolor[HTML]{F7FBFF} \color[HTML]{000000} 2,231.98} & {\cellcolor[HTML]{F7FBFF} \color[HTML]{000000} 1,670.03} & {\cellcolor[HTML]{F7FBFF} \color[HTML]{000000} 1,140.49} & {\cellcolor[HTML]{F7FBFF} \color[HTML]{000000} 3,776.17} & {\cellcolor[HTML]{F7FBFF} \color[HTML]{000000} 2,867.03} & {\cellcolor[HTML]{F7FBFF} \color[HTML]{000000} 2,001.03} \\
ID1 & {\cellcolor[HTML]{08306B} \color[HTML]{F1F1F1} 1,937.37} & {\cellcolor[HTML]{08306B} \color[HTML]{F1F1F1} 1,272.97} & {\cellcolor[HTML]{08306B} \color[HTML]{F1F1F1} 776.52} & {\cellcolor[HTML]{08306B} \color[HTML]{F1F1F1} 4,417.39} & {\cellcolor[HTML]{08306B} \color[HTML]{F1F1F1} 2,967.35} & {\cellcolor[HTML]{08306B} \color[HTML]{F1F1F1} 1,864.08} & {\cellcolor[HTML]{08306B} \color[HTML]{F1F1F1} 8,525.10} & {\cellcolor[HTML]{08306B} \color[HTML]{F1F1F1} 5,763.10} & {\cellcolor[HTML]{08306B} \color[HTML]{F1F1F1} 3,581.07} & {\cellcolor[HTML]{08306B} \color[HTML]{F1F1F1} 16,126.05} & {\cellcolor[HTML]{08306B} \color[HTML]{F1F1F1} 10,951.41} & {\cellcolor[HTML]{08306B} \color[HTML]{F1F1F1} 6,942.16} \\
\bottomrule
\end{tabular}%
    }
    \caption{Profits in 1000 EUR for no trading delay by forecasting model, interval $\alpha$, storage size and duration. Data is 2023 and 2024.}
\end{table}

\begin{table}[htb]  
    \adjustbox{width=\textwidth}{
        \begin{tabular}{llcccccccccccc}
\toprule
 &  & \multicolumn{3}{r}{10 MWh} & \multicolumn{3}{r}{25 MWh} & \multicolumn{3}{r}{50 MWh} & \multicolumn{3}{r}{100 MWh} \\
 &  & h=1 & h=2 & h=4 & h=1 & h=2 & h=4 & h=1 & h=2 & h=4 & h=1 & h=2 & h=4 \\
\midrule
\multirow[c]{9}{*}{Naive Day-Ahead} & $\alpha$=10 & {\cellcolor[HTML]{F7FCF5}} \color[HTML]{000000} 283.24 & {\cellcolor[HTML]{F7FCF5}} \color[HTML]{000000} 411.95 & {\cellcolor[HTML]{F7FCF5}} \color[HTML]{000000} 395.85 & {\cellcolor[HTML]{F7FCF5}} \color[HTML]{000000} 569.30 & {\cellcolor[HTML]{F7FCF5}} \color[HTML]{000000} 985.21 & {\cellcolor[HTML]{F7FCF5}} \color[HTML]{000000} 974.41 & {\cellcolor[HTML]{F7FCF5}} \color[HTML]{000000} 978.19 & {\cellcolor[HTML]{F7FCF5}} \color[HTML]{000000} 1840.57 & {\cellcolor[HTML]{F7FCF5}} \color[HTML]{000000} 1903.80 & {\cellcolor[HTML]{F7FCF5}} \color[HTML]{000000} 1620.62 & {\cellcolor[HTML]{F7FCF5}} \color[HTML]{000000} 3464.26 & {\cellcolor[HTML]{F7FCF5}} \color[HTML]{000000} 3670.92 \\
 & $\alpha$=20 & {\cellcolor[HTML]{E8F6E4}} \color[HTML]{000000} 320.23 & {\cellcolor[HTML]{F0F9ED}} \color[HTML]{000000} 423.22 & {\cellcolor[HTML]{EDF8EA}} \color[HTML]{000000} 403.29 & {\cellcolor[HTML]{F0F9ED}} \color[HTML]{000000} 612.78 & {\cellcolor[HTML]{F2FAF0}} \color[HTML]{000000} 1003.52 & {\cellcolor[HTML]{EFF9EC}} \color[HTML]{000000} 988.74 & {\cellcolor[HTML]{E5F5E1}} \color[HTML]{000000} 1188.02 & {\cellcolor[HTML]{EDF8EA}} \color[HTML]{000000} 1914.04 & {\cellcolor[HTML]{F0F9ED}} \color[HTML]{000000} 1929.27 & {\cellcolor[HTML]{E9F7E5}} \color[HTML]{000000} 1962.87 & {\cellcolor[HTML]{EEF8EA}} \color[HTML]{000000} 3597.63 & {\cellcolor[HTML]{F0F9EC}} \color[HTML]{000000} 3721.60 \\
\bottomrule
\end{tabular}

    } \\[1em]
    \adjustbox{width=\textwidth}{
        \begin{tabular}{llcccccccccccc}
\toprule
 &  & \multicolumn{3}{r}{10 MWh} & \multicolumn{3}{r}{25 MWh} & \multicolumn{3}{r}{50 MWh} & \multicolumn{3}{r}{100 MWh} \\
 &  & h=1 & h=2 & h=4 & h=1 & h=2 & h=4 & h=1 & h=2 & h=4 & h=1 & h=2 & h=4 \\
\midrule
\multirow[c]{9}{*}{Perfect Foresight} & $\alpha$=10 & {\cellcolor[HTML]{00441B}} \color[HTML]{F1F1F1} 3096.87 & {\cellcolor[HTML]{00441B}} \color[HTML]{F1F1F1} 1974.61 & {\cellcolor[HTML]{00441B}} \color[HTML]{F1F1F1} 1172.70 & {\cellcolor[HTML]{00441B}} \color[HTML]{F1F1F1} 7071.31 & {\cellcolor[HTML]{00441B}} \color[HTML]{F1F1F1} 4658.90 & {\cellcolor[HTML]{00441B}} \color[HTML]{F1F1F1} 2845.32 & {\cellcolor[HTML]{00441B}} \color[HTML]{F1F1F1} 13201.53 & {\cellcolor[HTML]{00441B}} \color[HTML]{F1F1F1} 8899.22 & {\cellcolor[HTML]{00441B}} \color[HTML]{F1F1F1} 5492.71 & {\cellcolor[HTML]{005120}} \color[HTML]{F1F1F1} 23968.92 & {\cellcolor[HTML]{00441B}} \color[HTML]{F1F1F1} 16857.02 & {\cellcolor[HTML]{00441B}} \color[HTML]{F1F1F1} 10661.60 \\
 & $\alpha$=20 & {\cellcolor[HTML]{006529}} \color[HTML]{F1F1F1} 2973.77 & {\cellcolor[HTML]{006D2C}} \color[HTML]{F1F1F1} 1883.86 & {\cellcolor[HTML]{00692A}} \color[HTML]{F1F1F1} 1127.97 & {\cellcolor[HTML]{005B25}} \color[HTML]{F1F1F1} 6887.79 & {\cellcolor[HTML]{00682A}} \color[HTML]{F1F1F1} 4481.45 & {\cellcolor[HTML]{006729}} \color[HTML]{F1F1F1} 2744.59 & {\cellcolor[HTML]{004C1E}} \color[HTML]{F1F1F1} 13077.06 & {\cellcolor[HTML]{005924}} \color[HTML]{F1F1F1} 8704.57 & {\cellcolor[HTML]{006428}} \color[HTML]{F1F1F1} 5320.49 & {\cellcolor[HTML]{00441B}} \color[HTML]{F1F1F1} 24284.65 & {\cellcolor[HTML]{005622}} \color[HTML]{F1F1F1} 16532.68 & {\cellcolor[HTML]{006027}} \color[HTML]{F1F1F1} 10376.23 \\
\bottomrule
\end{tabular}

    }
    \caption{Profits in 1000 EUR by forecasting model, interval $\alpha$, storage size and duration. Data is 2023 and 2024.}
\end{table}

\begin{table}[htb]
    \adjustbox{width=\textwidth}{
        \begin{tabular}{llcccccccccccc}
\toprule
 &  & \multicolumn{3}{r}{10 MWh} & \multicolumn{3}{r}{25 MWh} & \multicolumn{3}{r}{50 MWh} & \multicolumn{3}{r}{100 MWh} \\
 &  & h=1 & h=2 & h=4 & h=1 & h=2 & h=4 & h=1 & h=2 & h=4 & h=1 & h=2 & h=4 \\
\midrule
\multirow[c]{9}{*}{Naive Day-Ahead} & $\alpha$=10 & {\cellcolor[HTML]{F7FCF5}} \color[HTML]{000000} 28.32 & {\cellcolor[HTML]{F7FCF5}} \color[HTML]{000000} 41.20 & {\cellcolor[HTML]{F7FCF5}} \color[HTML]{000000} 39.59 & {\cellcolor[HTML]{F7FCF5}} \color[HTML]{000000} 22.77 & {\cellcolor[HTML]{F7FCF5}} \color[HTML]{000000} 39.41 & {\cellcolor[HTML]{F7FCF5}} \color[HTML]{000000} 38.98 & {\cellcolor[HTML]{F7FCF5}} \color[HTML]{000000} 19.56 & {\cellcolor[HTML]{F7FCF5}} \color[HTML]{000000} 36.81 & {\cellcolor[HTML]{F7FCF5}} \color[HTML]{000000} 38.08 & {\cellcolor[HTML]{F7FCF5}} \color[HTML]{000000} 16.21 & {\cellcolor[HTML]{F7FCF5}} \color[HTML]{000000} 34.64 & {\cellcolor[HTML]{F7FCF5}} \color[HTML]{000000} 36.71 \\
 & $\alpha$=20 & {\cellcolor[HTML]{E8F6E4}} \color[HTML]{000000} 32.02 & {\cellcolor[HTML]{F0F9ED}} \color[HTML]{000000} 42.32 & {\cellcolor[HTML]{EDF8EA}} \color[HTML]{000000} 40.33 & {\cellcolor[HTML]{F0F9ED}} \color[HTML]{000000} 24.51 & {\cellcolor[HTML]{F2FAF0}} \color[HTML]{000000} 40.14 & {\cellcolor[HTML]{EFF9EC}} \color[HTML]{000000} 39.55 & {\cellcolor[HTML]{E5F5E1}} \color[HTML]{000000} 23.76 & {\cellcolor[HTML]{EDF8EA}} \color[HTML]{000000} 38.28 & {\cellcolor[HTML]{F0F9ED}} \color[HTML]{000000} 38.59 & {\cellcolor[HTML]{E9F7E5}} \color[HTML]{000000} 19.63 & {\cellcolor[HTML]{EEF8EA}} \color[HTML]{000000} 35.98 & {\cellcolor[HTML]{F0F9EC}} \color[HTML]{000000} 37.22 \\
\bottomrule
\end{tabular}

    } \\[1em]
    \adjustbox{width=\textwidth}{
        \begin{tabular}{llcccccccccccc}
\toprule
 &  & \multicolumn{3}{r}{10 MWh} & \multicolumn{3}{r}{25 MWh} & \multicolumn{3}{r}{50 MWh} & \multicolumn{3}{r}{100 MWh} \\
 &  & h=1 & h=2 & h=4 & h=1 & h=2 & h=4 & h=1 & h=2 & h=4 & h=1 & h=2 & h=4 \\
\midrule
\multirow[c]{9}{*}{Perfect Foresight} & $\alpha$=10 & {\cellcolor[HTML]{00441B}} \color[HTML]{F1F1F1} 309.69 & {\cellcolor[HTML]{00441B}} \color[HTML]{F1F1F1} 197.46 & {\cellcolor[HTML]{00441B}} \color[HTML]{F1F1F1} 117.27 & {\cellcolor[HTML]{00441B}} \color[HTML]{F1F1F1} 282.85 & {\cellcolor[HTML]{00441B}} \color[HTML]{F1F1F1} 186.36 & {\cellcolor[HTML]{00441B}} \color[HTML]{F1F1F1} 113.81 & {\cellcolor[HTML]{00441B}} \color[HTML]{F1F1F1} 264.03 & {\cellcolor[HTML]{00441B}} \color[HTML]{F1F1F1} 177.98 & {\cellcolor[HTML]{00441B}} \color[HTML]{F1F1F1} 109.85 & {\cellcolor[HTML]{005120}} \color[HTML]{F1F1F1} 239.69 & {\cellcolor[HTML]{00441B}} \color[HTML]{F1F1F1} 168.57 & {\cellcolor[HTML]{00441B}} \color[HTML]{F1F1F1} 106.62 \\
 & $\alpha$=20 & {\cellcolor[HTML]{006529}} \color[HTML]{F1F1F1} 297.38 & {\cellcolor[HTML]{006D2C}} \color[HTML]{F1F1F1} 188.39 & {\cellcolor[HTML]{00692A}} \color[HTML]{F1F1F1} 112.80 & {\cellcolor[HTML]{005B25}} \color[HTML]{F1F1F1} 275.51 & {\cellcolor[HTML]{00682A}} \color[HTML]{F1F1F1} 179.26 & {\cellcolor[HTML]{006729}} \color[HTML]{F1F1F1} 109.78 & {\cellcolor[HTML]{004C1E}} \color[HTML]{F1F1F1} 261.54 & {\cellcolor[HTML]{005924}} \color[HTML]{F1F1F1} 174.09 & {\cellcolor[HTML]{006428}} \color[HTML]{F1F1F1} 106.41 & {\cellcolor[HTML]{00441B}} \color[HTML]{F1F1F1} 242.85 & {\cellcolor[HTML]{005622}} \color[HTML]{F1F1F1} 165.33 & {\cellcolor[HTML]{006027}} \color[HTML]{F1F1F1} 103.76 \\
\bottomrule
\end{tabular}

    }
    \caption{Profits in 1000 EUR by forecasting model, interval $\alpha$, storage size and duration. Normalized by the installed storage capacity. Data is 2023 and 2024.}
\end{table}

\end{document}